\documentclass[12pt]{article}
\usepackage{epsfig,rotating}
\def\lsim{\raise0.3ex\hbox{$\;<$\kern-0.75em\raise-1.1ex\hbox{$\sim\;$}}}
\def\gsim{\raise0.3ex\hbox{$\;>$\kern-0.75em\raise-1.1ex\hbox{$\sim\;$}}}
\begin{document}
\title{{\bf Analytic calculations of the spectra of ultra high energy 
cosmic ray nuclei.\\
II. The general case of background radiation.}}
   
\author{R. Aloisio$^{1,2}$, V. Berezinsky$^{2}$ and S. Grigorieva$^{3}$\\
        {\it\small $^1$INAF, Osservatorio Astrofisico Arcetri, 
        I--50125 Arcetri (FI), Italy} \\
        {\it\small $^2$INFN, Laboratori Nazionali del Gran Sasso,
        I--67010 Assergi (AQ), Italy} \\
        {\it\small $^3$Institute for Nuclear Research, 60th October 
        Revolution Prospect 7A, Moscow, Russia}
        }

\date{\today}
\maketitle

\abstract{ We discuss the problem of ultra high energy nuclei propagation in 
extragalactic background radiations. The present paper is the continuation 
of the accompanying paper I where we have presented three new analytic 
methods to calculate the fluxes and spectra of ultra high energy cosmic 
ray nuclei, both primary and secondary, and secondary protons. 
The computation scheme in this paper is based on the analytic 
solution of coupled kinetic equations, which takes into account the 
continuous energy losses due to the expansion of the universe and
pair-production, together with photo-disintegration of nuclei. 
This method includes in the most natural way the production of secondary
nuclei in the process of photo-disintegration of the primary nuclei
during their propagation through extragalactic background radiations. 
In paper I, in order to present the suggested analytical schemes of 
calculations, we have considered only the case of the cosmic microwave 
background radiation, in the present paper we generalize this 
computation to all relevant background radiations, including infra-red 
and visible/ultra-violet radiations, collectively referred to as extragalactic 
background light. The analytic solutions allow transparent physical 
interpretation of the obtained spectra. Extragalactic background light
plays an important role at intermediate energies of ultra high energy cosmic 
ray nuclei. The most noticeable effect of the extragalactic background light
is the low-energy tail in the spectrum of secondary nuclei.
}
\newpage

\section{Introduction}
\label{sec:intro}

In the accompanying paper \cite{paper1}, hereafter paper I, we have 
discussed three different analytic methods to study the propagation of 
Ultra High Energy Cosmic Ray (UHECR) nuclei through background 
radiations. In order to give a clear explanation of the analytic procedures, 
we have only discussed propagation in the Cosmic Microwave Background 
(CMB) neglecting all other relevant backgrounds. In the present paper we 
will extend our method including all relevant background radiations, obtaining
more complete results for the expected UHE nuclei spectra. Apart 
from the CMB we include the interaction with  Infrared (IR) and optical
photons, to which we refer collectively as Extragalactic Background Light 
(EBL).

The importance to study the UHECR nuclei as primary radiation have been
already discussed in the Introduction of paper I. It is enough to remind here 
again that according to recent Auger data \cite{Auger-mass}, the 
primary UHECR at energy higher than $(3 - 4)\times 10^{18}$~ eV are 
dominated by heavy nuclei. Propagation of nuclei through extragalactic 
background radiation results in a distortion of their energy spectra, different 
from that for UHE protons. There are many papers, cited in paper I, where 
propagation of UHE nuclei through CMB and EBL have been studied by 
Monte Carlo (MC) simulations. One of the most detailed study has been performed 
in \cite{allard08}. In the present paper we study this propagation 
analytically using the Coupled Kinetic Equations (CKE method, see paper I),
with both CMB and EBL radiations being included. We compute 
the fluxes and spectra of UHECR nuclei (primary and secondary) 
and secondary protons using, as in paper I, the hypothesis of a power-law 
generation spectrum, and assuming that source composition is given by 
nuclei with fixed atomic mass number $A_0$. We focus in this paper on the 
influence of the EBL on the propagation of UHE nuclei, discussing the effects 
of this background on the predicted spectra in comparison with the CMB. 

The EBL radiation is emitted by astrophysical objects 
at present and past cosmological epochs and subsequently is modified by 
red-shift and dilution due to the expansion of the Universe. 
The EBL energy spectrum is dominated by  
two peaks one at the optical and the other at IR energies, 
produced respectively by direct emission from
stars, and by thermal radiation from dust.  
At lower energies the  background spectrum is completely 
dominated by the CMB. 

Measurements of the EBL using direct observations is very difficult 
because of the foreground emissions, mainly
from our own galaxy, interplanetary dust radiation and reflected 
zodiacal light from the Sun \cite{hauser}. 
The optical EBL flux was evaluated by the measurements of
the wide-field planetary camera on board of the Hubble space telescope. 
The procedure consists in measuring the total background in three
different bands and subtracting the zodiacal light and the
galactic  foregrounds \cite{Bernstein}. The near-IR
flux was measured by the DIRBE instrument onboard the Cosmic 
Background Explorer (COBE) satellite
\cite{DIRBE1}, also such observations are affected by source 
subtraction techniques and modeling of the
zodiacal light. In the far-IR regime the EBL can be directly 
observed with less pollution from foregrounds, these
observations have been carried out by DIRBE \cite{DIRBE2} and 
FIRAS \cite{FIRAS} instruments both onboard COBE. 

Indirect observations of EBL are also used \cite{hauser}. 
One indirect method is based on the
integration of galaxy counts that helps in setting the reliable lower 
limits to the expected background and also in
determining the spectral energy distribution of the EBL, 
mainly at frequencies for which no COBE data
are available.  Another indirect way of evaluating the EBL density 
is based on the observations of TeV $\gamma$-rays \cite{hauser}, 
using the pair-production absorption 
features of $\gamma\gamma_{EBL}\to e^{+}e^{-}$. Using it  
one can deduce the intensity of EBL. Using 
TeV $\gamma$-rays observations
from blazars, the upper limits on the expected EBL have been obtained 
(see \cite{hauser} and references therein). 

For calculation of UHE nuclei spectra, the knowledge of EBL at early 
cosmological epochs is important and thus EBL cosmological evolution   
is needed. As discussed in \cite{kneiske} there were proposed three 
different methods to determine the EBL
cosmological evolution: (i) evolution inferred from 
observations at different red-shifts, (ii) forward evolution,
which begins with cosmological initial conditions and evolves them 
forward in time matching the present day
observations \cite{primack} and (iii) backward evolution, which starts 
form the present day observations and evolves
data backward in time \cite{stecker06}. At present there are a few 
works with calculations of the EBL with cosmological evolution
included, most notably \cite{stecker06} and \cite{kneiske04}. 
In the present paper we mainly use the EBL as presented in
\cite{stecker06}, which is a refinement of 
previous calculations \cite{malkan}, based on the data from the 
Spitzer infrared observatory and the Hubble 
Space Telescope deep survey. In \cite{stecker06} the EBL  photon 
density is found from $0.03$ eV up to
the Lyman limit $13.6$ eV for different values of the red-shift up to $z=6$. 

The paper is organized as follows: in section \ref{sec:losses} we 
discuss the energy losses of 
nuclei, focusing mainly on the effects of the EBL, in section 
\ref{sec:CKE} we briefly review the CKE
method and include in calculation the photo-disintegration of 
nuclei with multiple-nucleon emission, in section \ref{sec:spectra} 
we present our results on the expected fluxes of primary and secondary 
nuclei and secondary nucleons, in section \ref{sec:comparison} the 
calculated spectra are compared with existing calculations, and  
finally in section \ref{sec:conclusions} the results are discussed.

\section{Nuclei energy losses and role of the EBL}
\label{sec:losses}

In this section we discuss the nuclei energy losses in presence of 
EBL in the general form valid for the trajectory methods and 
kinetic equations (see paper I). 

Propagation of UHE nuclei through background radiations
is affected by three kinds of energy losses: (i) adiabatic 
losses due to the expansion of the Universe, 
(ii) losses due to  $e^+e^-$-pair production on 
the background  photons (these two interactions conserve 
the nuclei specie, i.e. $A$ and $Z$) and (iii) photo-disintegration of 
UHE nuclei (this process 
changes the nuclei specie giving rise to the production of 
secondary nuclei and nucleons). Presence of the EBL high energy photons  
requires to include into consideration photo-disintegration with 
multiple-nucleon production.  

Using the same approach as in  paper I, we will consider here two basic 
quantities that characterize the propagating nucleus, namely its atomic 
mass number $A$ and the Lorentz factor  $\Gamma$. The use  of the Lorentz 
factor instead of energy is more suitable because in the process of 
photo-disintegration, e.g. $(A+i) \to A + iN$ ($i=1,~2,~3 ..$), the 
Lorentz-factors of all particles are approximately the same, since the 
kinetic recoil energy of the secondary nucleus is much smaller 
than the rest-mass of this particle. Therefore, during propagation the 
nucleus  Lorentz factor changes only due to the expansion of the universe and  
pair-production, remaining unchanged in the process of photo-disintegration.

In paper I we have introduced three different analytic schemes 
to compute the fluxes of UHECR nuclei and their secondaries. 
All these methods are based on the solution of kinetic equations.  
The photo-disintegration process is interpreted there as a decaying process, 
e.g. $A \to (A-1) +N$, which results in the disappearance
of the nucleus $A$. In this sense only the pair-production
process and the Universe expansion change the Lorentz factor 
$\Gamma$, and thus energy $E=\Gamma A m_N$, until the 
disappearance of the nucleus $A$. The photo-disintegration of $A$ 
nuclei is the only process which depletes the $A$-nuclei flux.

The energy spectra of UHECR nuclei have been calculated  
in paper I assuming only CMB radiation. The calculations in 
this work differ by the presence of 
EBL radiation, which affects only the low energy part of the spectrum. 
Since energies of EBL photons are higher than that of CMB, the 
calculated spectra are characterised by photo-disintegration 
which occurs at lower
energies. One may  immediately understand that EBL radiation makes 
negligible contribution to pair-production at all energies. To explain it 
let us start with low Lorentz-factors $\Gamma$, when the pair
production occurs on optical and UV photons of the EBL spectrum. 
The threshold of pair production is given by 
$\Gamma_{\rm th} \epsilon \sim 2m_e$, where $\epsilon$  
is the energy of EBL photon. For $\epsilon \sim 1$~eV,~ 
$\Gamma_{\rm th} \sim 10^6$, and for known density of EBL 
photons the energy losses due to pair production are considerably lower 
than that for adiabatic energy loss $H_0$. At higher Lorentz factors 
the pair-production on CMB photons strongly dominates because of
much larger density of CMB photons in comparison with the EBL. The
numerical calculations confirm this conclusion.  

Let us now come over to the numerical discussion of energy losses 
and role of EBL and multiple-nucleon photo-disintegration. 

The rate of the Lorentz factor loss due to pair-production, i.e. energy 
loss for fixed $A$, can be written for all Lorentz factors $\Gamma$ 
taking into account only CMB radiation. It easily can be written in 
terms of pair-production process for protons (see also 
section 2 of paper I) as  

\begin{equation}
\left ( \frac{1}{\Gamma}\frac{d\Gamma}{dt} \right)_{pair}^A \equiv 
\beta^A_{\rm pair}(\Gamma,t) = \frac{Z^2}{A}\beta^{p}_{\rm pair}(\Gamma,t)
\label{eq:pair}
\end{equation} 
where $Z$ and $A$ are, respectively, the electric-charge number and 
atomic-mass number of the nucleus, 
and $\beta^{p}_{\rm pair}(\Gamma,t)$ is the Lorentz factor decrease rate 
for the proton on CMB \cite{bere-gaz-grig06}.  

\begin{figure}[t!]
\begin{center}
\includegraphics[width=0.9\textwidth,angle=0]{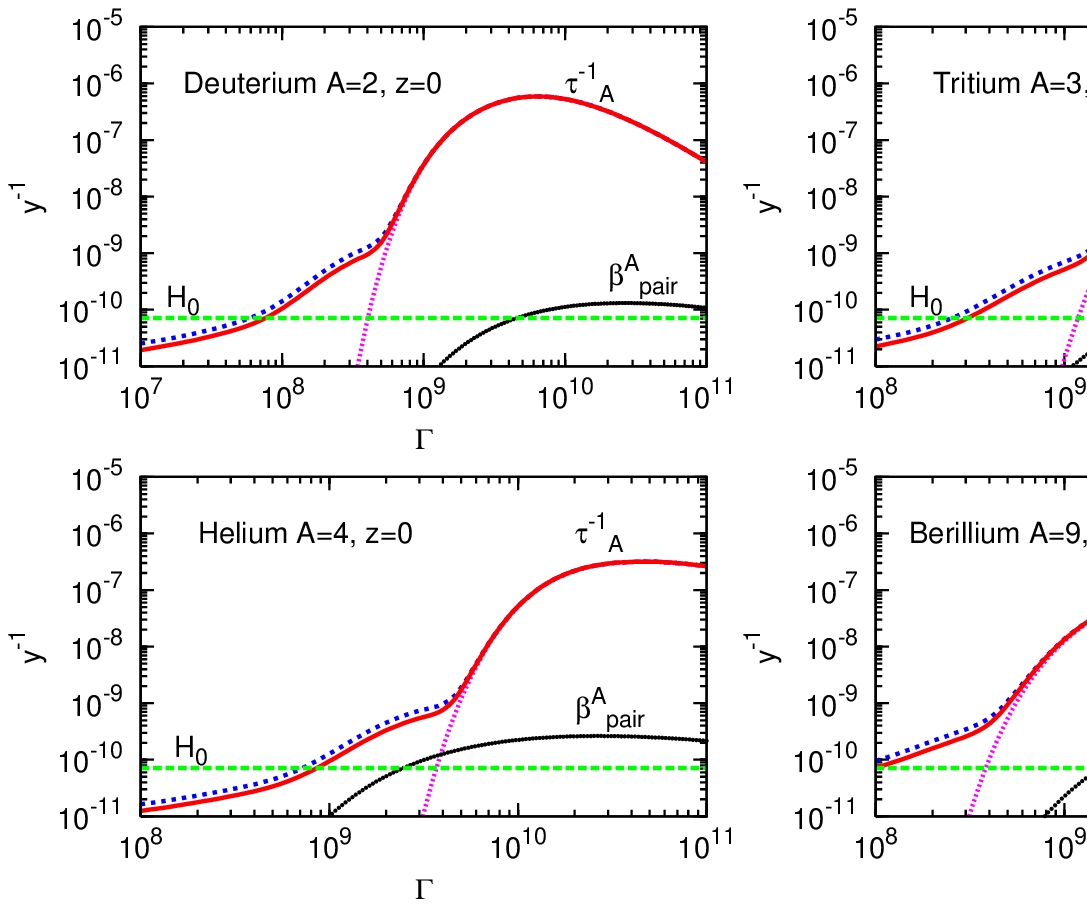}
\caption{Photo-disintegration lifetime plotted as $\tau^{-1}_{A}$, 
for light nuclei at $z=0$ as a function of the nucleus Lorentz factor 
$\Gamma$ for the case of CMB alone (magenta dotted) 
and for two cases of EBL: baseline (red continuous) and fast evolution  
(blue dotted). Lorentz factor decrease rate $\beta$ due to pair
production (black continuous), is given here only for CMB (EBL 
contribution is negligible). The Hubble constant $H_0$ at $z=0$ 
(green dashed) gives the adiabatic energy losses due to the 
expansion of the Universe. Note that the fraction of kinetic energy
lost is given by $\beta_{\rm pair}\tau_A$ and $H_0\tau_A$.
}
\label{fig1}
\end{center}
\end{figure}

\begin{figure}[t!]
\begin{center}
\includegraphics[width=0.9\textwidth,angle=0]{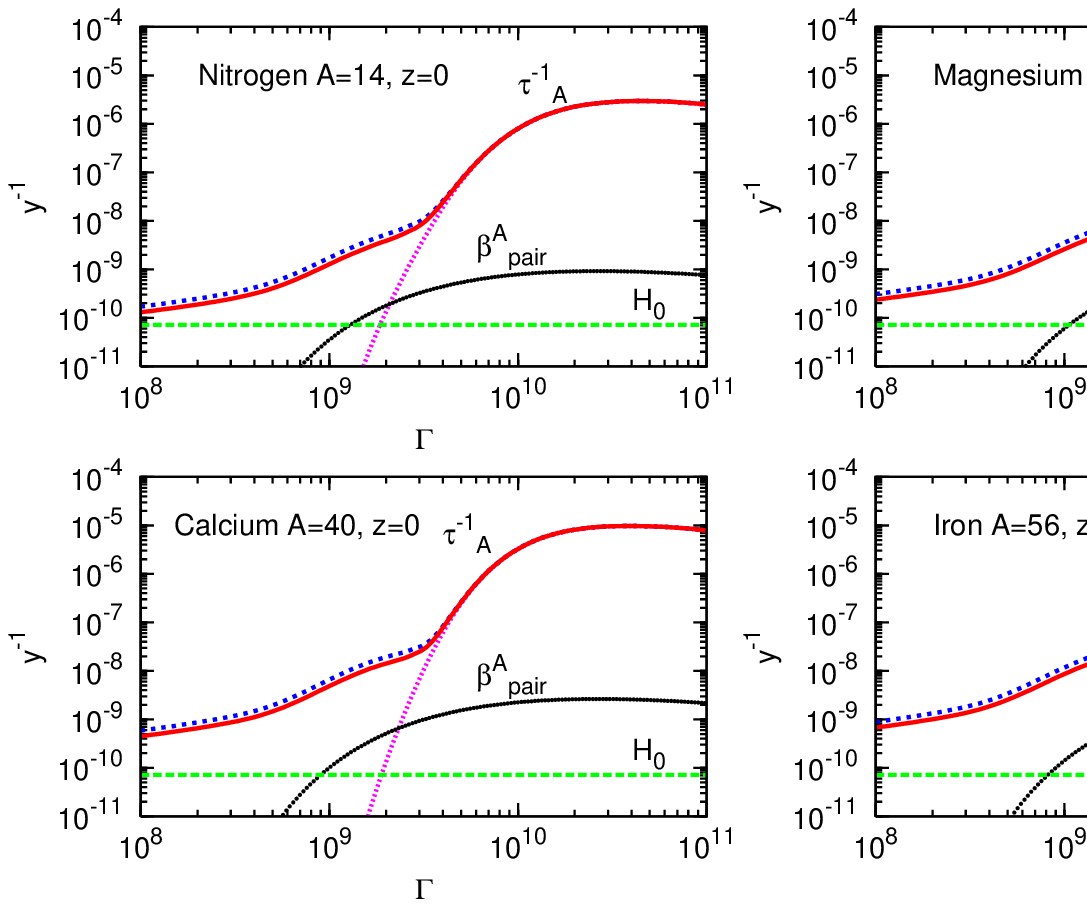}
\caption{The same as in Fig.~\ref{fig1} for heavy nuclei.}
\label{fig2}
\end{center}
\end{figure}

The effect of EBL is relevant only for the process of 
photo-disintegration of nuclei. This process can be described 
with the help of the quantity $dA/dt$ determined as  
\begin{equation}
\frac{dA}{dt}=\frac{c}{2\Gamma^2}
\int_{\epsilon_0(A)}^{\infty} d\epsilon_r \sigma(\epsilon_r,A)\nu(\epsilon_r)\epsilon_r
\int_{\epsilon_r/(2\Gamma)}^{\infty} d\epsilon \frac{n_{bcgr}(\epsilon)}{\epsilon^2},
\label{eq:tauA}
\end{equation}
where $\epsilon$ and $\epsilon_r$ are the energies of background photons in 
the laboratory system and in the rest system of the nucleus, respectively,
$n_{bcgr}=(n_{CMB}+n_{EBL})$ is the photon density of  
CMB and EBL background radiations,
$\sigma$ and $\nu$ are, respectively, the photo-disintegration 
cross-section and the multiplicity (mean number) of
ejected nucleons (due to presense of EBL we do not assume here
$\nu=1$). For the cross-sections with different multiplicities  
we use the cross section parameterization from 
\cite{stecker-photo1,stecker-photo2}.

One may define the characteristic time $\tau_A$ from the relation 
$(dA/dt)\tau_A=1$, which has the meaning of the average time needed 
for a nucleus $A$ to lose one nucleon in the interaction with 
background photons. Note that the fraction of kinetic energy 
lost by a nucleus A during its lifetime is given by 
$\beta_{pair}^A \tau_A$. 

In the case of CMB alone, discussed in paper I, the evolution of 
$\tau_A$ with red-shift (hereafter we will refer 
to red-shift instead of cosmological time) was simply fixed by the 
CMB evolution: the number of CMB photons
increases by a factor $(1+z)^3$ and their energy by a factor 
$(1+z)$. As discussed in the Introduction the EBL evolution with 
red-shift is not reliably known and 
several models have been put forward to describe 
such evolution. In the present paper we will use the 
evolution model of Stecker et al.
\cite{stecker06}. In this model the dependence of the EBL spectral 
distribution on red-shift is determined 
through a backward evolution in time of the observed spectral 
distribution at $z=0$. This method is an empirically
based calculation of the spectral energy distribution  of EBL 
using: (i) the luminosity dependent spectral energy distribution
of galaxies based on the observations of normal galaxies, 
(ii) observationally based luminosity functions as 
discussed in \cite{sauders90} and (iii) the red-shift dependent 
luminosity evolution functions, empirically 
derived curves giving the universal star formation rate 
\cite{madau96} or luminosity density \cite{lilly96}.
The calculations of the work \cite{stecker06} are based on two different 
scenarios for the luminosity evolution: the 
base-line scenario and the fast evolution scenario. 

In the base-line model galactic luminosities at 60 $\mu$m evolve 
as $(1+z)^{3.1}$ up to
$z=1.4$, at higher red-shifts the luminosity is assumed constant 
with negligible emission at red-shift $z>6$. In
particular, this last assumption of the base-line scenario is supported 
by the observations of the Hubble space
telescope, which indicate that the star formation rate drops off 
significantly at red-shift around $z=6$
\cite{bouwens06}, similar decrease is also reported by the 
Subaru deep field observations of the Ly$\alpha$
emitting objects at red-shift $z=6.5$ \cite{kashikawa06}. 

In the fast evolution scenario the galaxies luminosity is evolved 
as $(1+z)^4$ in the red-shift range 
$0<z<0.8$ and as $(1+z)^2$ in the range $0.8<z<1.5$. At higher 
red-shifts all luminosities are assumed 
constant with no evolution and, as in the baseline scenario, the 
luminosity is assumed zero at $z>6$. This
kind of evolution is based on the mid-IR luminosity functions 
determined from that at $z=2$ in \cite{perez05}. 

\begin{figure}[t!]
\begin{center}
\includegraphics[width=0.9\textwidth,angle=0]{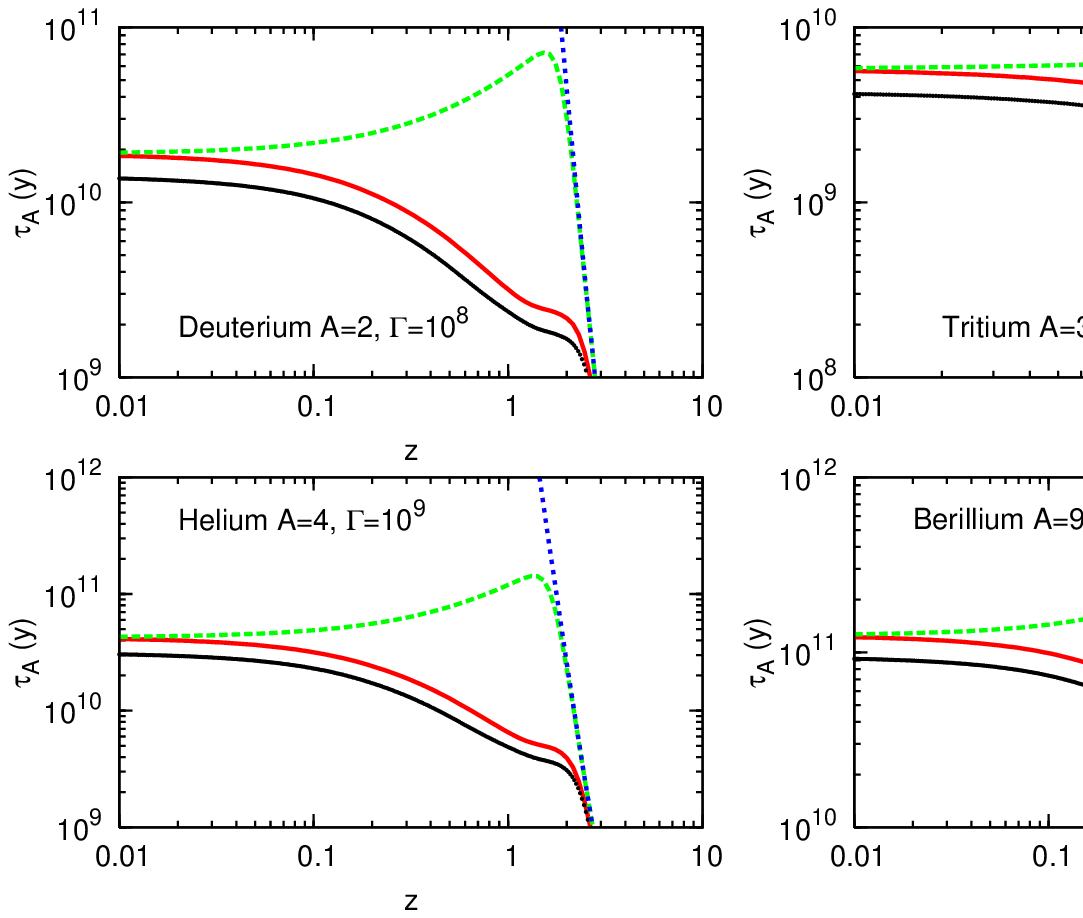}
\caption{Photo-disintegration life-time $\tau_{A}$ for light nuclei 
as a function of the red-shift for the fixed value of the 
nucleus Lorentz factor. Three different models of the
EBL evolution are used: baseline model (red continuous), fast evolution 
model (black continuous) and minimum EBL (green dashed).
Blue dotted curve presents the case when only the CMB is taken into
account.
}
\label{fig3}
\end{center}
\end{figure}

\begin{figure}[t!]
\begin{center}
\includegraphics[width=0.9\textwidth,angle=0]{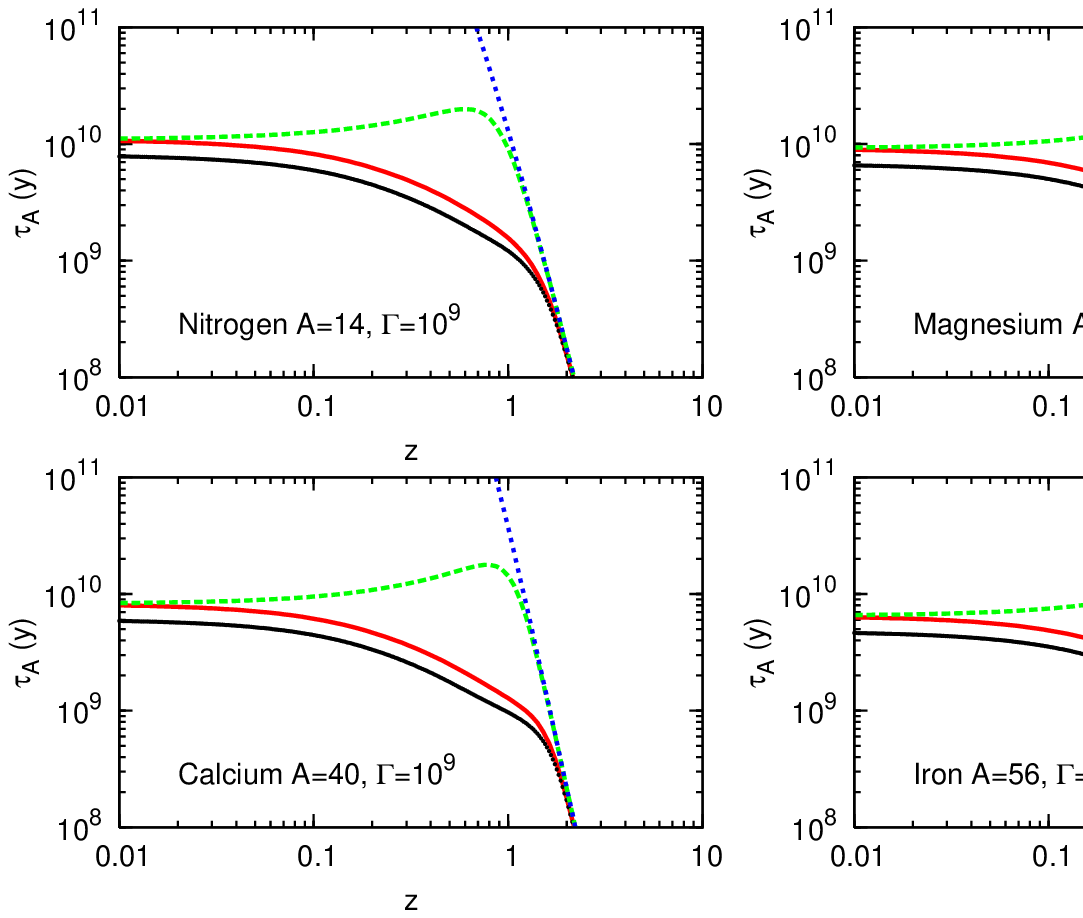}
\caption{The same as in figure \ref{fig3} for heavy nuclei.}
\label{fig4}
\end{center}
\end{figure}

The fast evolution model corresponds to somewhat like upper limit for 
the EBL density, with the larger 
contribution at high red-shift. As a lower limit to the EBL 
contribution we have also considered a third 
possibility that corresponds to the minimum possible EBL density 
at $z>0$. This density can be found 
from the following general statement, which we have proved for any 
diffuse background radiation: 
In the case of the generation rate of background radiation 
$Q(\epsilon,z)=K\epsilon^{-\alpha}(1+z)^m$, 
valid  up to $z_{\rm max}$, with arbitrary $\alpha$, 
$m \geq 0$ and assuming that the background photons are
not absorbed, the density of diffuse background radiation at epoch 
$z$ is always larger than
$n_z(\epsilon)=(1+z)^{-3/2} n_0(\epsilon)$, where $n_0(\epsilon)$ 
is the measured density at $z=0$. 

Once the dependence of the EBL photon density on the red-shift is 
determined, one can write explicitly the 
photo-disintegration "life-time" for a nucleus $A$ with Lorentz 
factor $\Gamma$ at any red-shift $z$ .
Separating the contributions from the two backgrounds, CMB and EBL, 
in equation (\ref{eq:tauA}), one has:
\begin{equation}
\frac{1}{\tau_A(\Gamma,z)}=\frac{1}{\tau^A_{CMB}(\Gamma,z)} + \frac{1}{\tau^A_{EBL}(\Gamma,z)}=
\label{eq:tauA2}
\end{equation}
$$=\frac{cT (1+z)}{2\pi^2\Gamma^2}
\int_{\epsilon_0(A)}^{\infty}d\epsilon_r\sigma(\epsilon_r,A)
\nu(\epsilon_r)\epsilon_r
\left [-\ln\left (1-e^{\frac{\epsilon_r}{2\Gamma (1+z) kT}}\right)\right ] +$$ 
$$+\frac{c}{2\Gamma^2}
\int_{\epsilon_0(A)}^{\infty} d\epsilon_r \sigma(\epsilon_r,A)\nu(\epsilon_r)\epsilon_r
\int_{\epsilon_r/(2\Gamma)}^{\infty} d\epsilon \frac{n_{EBL}
(\epsilon,z)}{\epsilon^2} $$
where $\tau^A_{CMB}(\Gamma,z)$ is the CMB contribution calculated in 
paper I and $\tau^A_{EBL}(\Gamma,z)$ is the EBL contribution, with 
$n_{EBL}(\epsilon,z)$ computed within the three evolutionary
models: the baseline and fast evolution models of \cite{stecker06},
and minimum EBL model, as described in this section. 

In Figs.~\ref{fig1} and \ref{fig2} we plot $\tau_A^{-1}$ and 
$\beta^{A}_{pair}$ at $z=0$ as function of $\Gamma$
for various nuclei species as labelled. We have plotted $\tau_A^{-1}$
for two different models of EBL evolution \cite{stecker06}, baseline 
(red continuous) and fast evolution (blue dotted), compared to the case  
of CMB alone (magenta dotted). The effect of EBL 
is clearly seen at
intermediate energies with a tiny difference among the two choices 
of baseline and fast evolution. 

To illustrate the effect of different evolution regimes we have
plotted in Figs.~\ref{fig3} and \ref{fig4}
the photo-disintegration life-time $\tau_A$ as function 
of the red-shift for two fixed values of the
nucleus Lorentz factors $\Gamma=1\times 10^8$ and 
$\Gamma=1\times 10^9$. This choice is motivated by Figs.~\ref{fig1} 
and \ref{fig2}, which show that the EBL effect is dominant 
in the range $10^{8}<\Gamma< 2\times 10^{9}$. Figures \ref{fig3} 
and \ref{fig4} show the variation of $\tau_A$ with $z$ for the three
regimes of EBL evolution discussed above. The red continuous 
curve corresponds to the baseline model, the
black continuous curve to the fast evolution scenario and the green dashed 
line to the minimum EBL, normalized at $z=0$ to the
baseline density. The effect of the EBL reveals itself  
with a longer life-time in the case of minimum EBL and
shorter in the case of fast evolution. This result can be easily 
understood by taking into account that when the EBL photon density
increases, the photo-disintegration process becomes more efficient and 
the corresponding nucleus life-time
decreases. The plots of figures \ref{fig3} and \ref{fig4} show also 
that the effect of the  EBL is efficient only at low red-shifts.
The evolution of the CMB is strong: the density of photons increases as 
$(1+z)^3$ and energies as $1+z$. Therefore, at large red-shifts the CMB
dominates. Figs.~\ref{fig3} and \ref{fig4} show that $\tau_A$ is
determined by the CMB already at $z \geq 2$. This evidence reduces 
the impact of the EBL evolution on our calculations: at $z < 2$ the
evolutionary models do not differ much and at $z >2$ CMB dominates.   

\section{Coupled Kinetic Equations (CKE)}
\label{sec:CKE}

In this section we develop further the CKE method, putting together 
all the details 
discussed in paper I. The modification consists in the inclusion of the 
EBL photon density 
$n_{\rm EBL}(\epsilon,z)$, which affects the photo-disintegration
lifetime $\tau_A(\Gamma,t)$ of a nucleus at cosmological epoch $t$. 

Another modification consists in the inclusion of multiple-nucleon emission 
in the process of photo-disintegration $(A+i)+ \gamma_{bcgr} \to A + iN$ 
with $i \geq 2$.
 
The basic kinetic equation for space density of $A$-nuclei 
$n_A(\Gamma,t)$ under the assumption of a homogeneous distribution of sources 
has the form: 
\begin{equation}
\frac{\partial n_{A}(\Gamma,t)}{\partial t} - \frac{\partial}{\partial \Gamma}
\left [ n_{A}(\Gamma,t) b_{A}(\Gamma,t) \right ] + \frac{n_{A}(\Gamma,t)}
{\tau_{A}^{\rm tot}(\Gamma,t)}  = Q_{A}(\Gamma,t)
\label{eq:kin1}
\end{equation}
where $t$ is the cosmological time, $b_A=-d\Gamma/dt$ is the rate of
Lorentz-factor loss, and $Q_{A}$ is the rate of $A$-nuclei production.  
Here and hereafter $\tau_A^{\rm tot}$ includes all channels of
photo-disintegration, with single and multiple nuclei emission. 
Eq.~(\ref{eq:kin1}) is valid for $A_0$, $A_0-1$ etc. 

In Eq.~(\ref{eq:kin1}) the rate $b_A$ of Lorentz-factor decrease includes 
the terms due to the expansion of the universe and pair-production on CMB, 
as discussed in the previous section, and using Eq.~(\ref{eq:pair}) for 
$\beta^A_{\rm pair}= b_{\rm pair}^A/\Gamma$ this rate can be explicitly written as 
\begin{equation}
b_A(\Gamma,z) = \Gamma\frac{Z^2}{A} \beta^p_{pair}(\Gamma,z) + 
\Gamma H_0\sqrt{\Omega_m (1+z)^3 +\Omega_{\Lambda}}.
\label{eq:bA}
\end{equation}

Like in paper I we consider here and hereafter the standard cosmology
with $H_0 \approx 72$~ km/(s Mpc), 
$\Omega_m \approx 0.27$ and $\Omega_\Lambda \approx 0.73$.
We will often use the Hubble parameter $H(z)$ at epoch $z$ and
$dt/dz$, given by 
\begin{equation}
 H(z)=H_0 \sqrt{(1+z)^3\Omega_m + \Omega_\Lambda}~~~{\rm and}~~~
\frac{dt}{dz}=- \frac{1}{(1+z)H(z)}.
\label{eq:H(z)}
\end{equation}

Let us discuss now the generation function $Q_A(\Gamma,t)$ in the rhs 
of Eq.~(\ref{eq:kin1}). For the primary nuclei $A_0$ this term
describes the injection rate of nuclei accelerated in the sources.  
Like in paper I we assume that only a single specie $A_0$ is
accelerated (in future applications one can sum up over different $A_0$).
As a particular example, in this paper we will consider as primary the Iron 
nuclei $A_0=56$. Assuming an homogeneous distribution of sources, the 
injection rate for these nuclei is given by   
\begin{equation}
Q_{A_0}(\Gamma,z)={\mathcal L}_0 \frac{(\gamma_g-2)}{m_N A_0} \Gamma^{-\gamma_g}
\label{eq:injA0}
\end{equation}
where $\gamma_g>2$ is the generation index, $m_N$ is the nucleon mass 
and ${\mathcal L}_0$ is the emissivity, i.e. the energy injected
per unit of comoving volume and per unit time.  

Propagating through background radiations Iron nuclei are
photo-disintegrated producing, through a photo-disintegration chain, 
secondary nuclei with $A < A_0$ and secondary nucleons 
(neutrons decay very fast to protons). Therefore, we
have only three types of propagating particles: primary nuclei $A_0$,
secondary nuclei with $A < A_0$ and secondary protons. 

The generation rate of secondary nuclei $A$ is determined by  
the photo-disintegration of heavier nuclei. 

The simplest and the dominant channel is given by single-nucleon 
photo-disintegration $(A+1) + \gamma_{bcgr} \to A+N$. All three
nuclear particles have the same Lorentz-factor $\Gamma$ and the
production rate of secondary $A$-nucleus and $A$-associate proton 
is given by 
\begin{equation}
Q_A(\Gamma,z)= Q_p^A(\Gamma,z)=
\frac{n_{A+1}(\Gamma,z)}{\tau_{A+1}^A(\Gamma,z)}
\label{eq:injA-single}
\end{equation}
Equation (\ref{eq:injA-single}) is the basis of the CKE method. The
generation rate of A-nuclei in Eq.~(\ref{eq:kin1}) is
determined by $n_{A+1}(\Gamma,z)$, which is found as solution 
of the preceding equation for $A+1$ nuclei. Therefore, 
Eq.~(\ref{eq:injA-single}) provides the coupling of two equations, for 
$A$ and $A+1$ nuclei (see section~4 of paper I for more  details).

The lifetime for $(A+1) \to A + N $ photo-disintegration, $\tau_{A+1}^A$,
is calculated as 
\begin{equation}
\left [\tau_{A+1}^A (\Gamma,z)\right ]^{-1} =\frac{c}{2\Gamma^2}
\int_{\epsilon_{\rm th}^{(1)}(A+1)}^{\infty} d\epsilon_r 
\sigma^{(1)}(\epsilon_r,A)\epsilon_r
\int_{\epsilon_r/(2\Gamma)}^{\infty} d\epsilon
\frac{n_{bcgr}(\epsilon)}{\epsilon^2},
\label{eq:tauA-single}
\end{equation}
where $\sigma^{(1)}$ and $\epsilon_{\rm th}^{(1)}$
are respectively cross-section and threshold associated to single-nucleon
photo-disintegration. 

For $i$-nucleons photo-disintegration 
$(A+i) + \gamma_{bcgr} \to A + iN$ with $i=2,~3 ...$ the generation
rates for $A$-nuclei 
and for each of the secondary protons are given by 
\begin{equation}
Q_A(\Gamma,z)= Q_p^A(\Gamma,z)=
\frac{n_{A+i}(\Gamma,z)}{\tau_{A+i}^A(\Gamma,z)},
\label{eq:injA-i}
\end{equation}
with 
\begin{equation}
\left [\tau_{A+i}^A (\Gamma,z)\right ]^{-1} =\frac{c}{2\Gamma^2}
\int_{\epsilon_{\rm th}^{(i)}(A+1)}^{\infty} d\epsilon_r 
\sigma^{(i)}(\epsilon_r,A)\epsilon_r
\int_{\epsilon_r/(2\Gamma)}^{\infty} d\epsilon
\frac{n_{bcgr}(\epsilon)}{\epsilon^2},
\label{eq:tauA-i}
\end{equation}

Finally, $\tau_A^{\rm tot}$ in Eq.~(\ref{eq:kin1}) is given by 
\begin{equation}
[\tau_A^{\rm tot}]^{-1}= \sum_i [\tau_A^{A-i}]^{-1}
\end{equation}

\subsection{CKE with single-nucleon emission}
\label{sec:onenucleon}

In the approximation of single-nucleon emission the generation
process for secondary A-nuclei production is  
$(A+1) +\gamma_{bcgr} \to A+N$. The production rate for 
secondary nuclei is given by Eq.~(\ref{eq:injA-single}) 
and lifetime - by Eq.~(\ref{eq:tauA-single}).
 
The first step in the chain of CKE is the equation for primaries $A_0$ 
with the generation term given by Eq.~(\ref{eq:injA0}): 
\begin{equation}
\frac{\partial n_{A_0}(\Gamma,t)}{\partial t} - 
\frac{\partial}{\partial \Gamma}
\left [ n_{A_0}(\Gamma,t) b_{A_0}(\Gamma,t) \right ] + \frac{n_{A_0}(\Gamma,t)}
{\tau_{A_0}^{\rm tot}(\Gamma,t)}  = Q_{A_0}(\Gamma,t).
\label{eq:kin0}
\end{equation}
Its solution reads 
\begin{equation}
n_{A_0}(\Gamma,z)=\int_{z}^{z_{max}} \frac{dz'}{(1+z') H(z')}
Q_{A_0}(\Gamma ',z')
\frac{d\Gamma '}{d\Gamma} e^{-\eta_{A_0}(\Gamma ',z')}. 
\label{eq:nA0-solut}
\end{equation}
The second chain is the equation (\ref{eq:kin1}) for $A_0-1$ nuclei with 
generation rate
$n_{A_0}(\Gamma',z')/\tau_{A_0}^{A_0-1}(\Gamma',z')$. 

For an arbitrary secondary nuclei $A$ the generation term
is provided by $n_{A+1}(\Gamma,z)$ found from the preceding equation,
and thus $n_A(\Gamma,z)$ is presented by
\begin{equation}
n_A(\Gamma,z)=\int_{z}^{z_{max}} \frac{dz'}{(1+z')H(z')} 
\frac{n_{A+1}(\Gamma',z')}{\tau_{A+1}^A(\Gamma',z')}
\frac{d\Gamma'}{d\Gamma} e^{-\eta_{A}(\Gamma',z')}.
\label{eq:nA-solut}
\end{equation}
The exponential term in Eq.~(\ref{eq:nA-solut}) is given by 
\begin{equation}
e^{-\eta_{A}(\Gamma ',z ')} = \exp\left [-\int_{z}^{z'} \frac{dz''}{(1+z'')H(z'')}
\frac{1}{\tau_{A}^{\rm tot}(\Gamma '', z'')}\right ] .
\label{eq:etaA}
\end{equation}
The physical meaning of the factor $\exp(-\eta)$ becomes clear if 
Eq.~(\ref{eq:etaA}) is re-written in terms of the cosmological time $t$ as  
\begin{equation}
e^{-\eta_{A}(\Gamma ',t ')} = \exp\left [-\int_{t}^{t'} \frac{dt''}
{\tau_{A}^{\rm tot}(\Gamma '', t'')}\right ],
\label{eq:etaAt}
\end{equation}
in which one easily recognizes the survival probability 
during the propagation time $t'-t$ for a nucleus with fixed $A$.
Therefore, $\exp(-\eta)$ provides the suppression of large $z'$ in 
the integral in Eq.~(\ref{eq:nA-solut}).

As we already emphasized, in each of the coupled kinetic equations 
$A=const$, though at any $z'$ a probability of $A=const$ is suppressed by 
$\exp(-\eta)$. Two consequences follow from such description. 

First, the maximum red-shift $z_{\max}$ in Eqs.~({\ref{eq:nA0-solut}) and 
(\ref{eq:nA-solut}) formally corresponds to the maximum acceleration
Lorentz factor $\Gamma_{\max}$ in the Lorentz factor evolution
at fixed $A$. As a matter of fact the suppression factor 
$\exp(-\eta)$ controls automatically the maximum attainable Lorentz factor. 
Therefore, one avoids using $z_{\min}$ and $z_{\max}$ from trajectory   
calculations. In fact $z_{\min}=z$, while effective $z_{\max}$ is controlled by 
survival probability $\exp(-\eta)$.

Second, ratios $d\Gamma/d\Gamma'$ in  Eqs.~({\ref{eq:nA0-solut}) and
(\ref{eq:nA-solut}) are characterised by  $A=const$, and therefore for
their calculation the formula (68) from appendix B of paper I is valid:
\begin{equation}
\frac{d\Gamma'}{d\Gamma}=\frac{1+z'}{1+z}
\exp\left [ \frac{Z^2}{A}\int_{z}^{z'} dz'' \frac{(1+z'')^2}{H(z'')} 
\left (\frac{d b_0^p (\tilde{\Gamma})}{d\tilde{\Gamma}} 
\right )_{\tilde{\Gamma}=(1+z'')\Gamma''} \right ] ,
\label{eq:ratio-dGamma}
\end{equation}
where $b_0^p(\Gamma)=- d\Gamma /dt$ is the Lorentz-factor loss per unit time 
for protons at $z=0$ due to pair production process. 

Finally, we address the calculation of the {\em secondary protons} associated
to the production of secondary $A$ nuclei in the process $(A+1)+\gamma \to A+N$. 
For the space density of $A$-associate protons the notation $n_p^A$
will be used.  

The kinetic equation for proton propagation reads 
\begin{equation}
\frac{\partial n_p(\Gamma,t)}{\partial t} - \frac{\partial}{\partial 
\Gamma} \left [ b_p(\Gamma,t)n_p(\Gamma,t) \right ] = Q_p(\Gamma,t)
\label{eq:kin_p}
\end{equation}   
where $Q_p$ is the rate of proton production given by Eq.~(\ref{eq:injA-single})
and $b_{p}(\Gamma,t)=-d\Gamma/dt$ is the Lorentz-factor decrease rate due 
to the expansion of the universe (adiabatic energy losses), 
pair-production and photo-pion production both 
on the CMB radiation field \cite{bere-gaz-grig06}.    

The solution of Eq.~(\ref{eq:kin_p}) is given by 
\begin{equation}
n^A_p(\Gamma,z)=\int_{z}^{z_{\rm max}} 
\frac{dz'}{(1+z)H(z)} Q^A_p(\Gamma',z')
\left (\frac{d\Gamma'}{d\Gamma}\right )_p
\label{eq:np}
\end{equation}
where $d\Gamma/d\Gamma'$ for protons is given in \cite{BG88} and paper
I. It can be calculated from Eq.~(\ref{eq:ratio-dGamma})
using the assumption relative to the {\it stability valley} for 
nuclei $Z=A/2$ ,
and including in $b_0^p(\Gamma)$ the photo-pion production energy losses.

\subsection{CKE with multiple-nucleon emission}
\label{sec:manynucleon}

In this subsection we include, additionally to the case of single-nucleon 
emission, multi-nucleon photo-disintegration. 

The numerical analysis of multiple-nucleon emission in the nuclei
photo-disintegration  is performed in 
\cite{stecker-photo1,stecker-photo2,AhTa2010}.
The main conclusion of these works is that multi-nucleon emission in the
production of secondary nuclei is a subdominant process. As was noticed   
first in \cite{stecker-photo2} the main effect in this phenomenon is provided
by higher energy threshold for multi-nuclei production. While a
typical energy threshold for a single-nucleon emission is 
$\epsilon_{\rm th} \sim 10 $~MeV for two-nucleon emission it is 
$\epsilon_{\rm th} \sim 20$~MeV (see Table~1 in \cite{stecker-photo2}). For 
CMB, the  suppression of secondary-nuclei flux with two-nucleon emission 
occurs due to the CMB-photon spectrum, while for EBL - due to the energy
spectrum of nuclei, and thus the former suppression is stronger than the 
latter. Production of alpha-particles is characterized by a low-energy
threshold, typically $\epsilon_{\rm th} \sim 7 - 10 $~MeV, but it is 
suppressed by smaller cross-sections. The physical quantity responsible for
the flux suppression is the photo-disintegration lifetime. Its increase with  
multiplicity growth is clearly seen in Fig.~2 of \cite{stecker-photo2}.   

In the present work we
include the terms with multiple-nucleon emission in A-nuclei 
generation rate. The relative smallness of this rate is provided
by larger lifetimes for multiple-nucleon photo-disintegration 
$(A+i) \rightarrow A+i N$ with $i\geq 2$ in comparison with the 
single-nucleon emission process $i=1$.

The generation rate and kinetic equation for {\em primary nuclei} $A_0$ remain 
the same and it is given by the equations (\ref{eq:injA0}) and (\ref{eq:kin0}). 

The kinetic equation for an arbitrary {\em secondary} nuclei $A$ is different 
from equation (\ref{eq:kin1}) and it reads 
\begin{equation}
\frac{\partial n_A(\Gamma,t)}{\partial t} - \frac{\partial}{\partial\Gamma}
\left [ n_A(\Gamma,t) b_A(\Gamma,t) \right ] + 
\frac{n_A(\Gamma,t)}{\tau_A^{\rm tot}(\Gamma,t)} = 
\frac{n_{A+1}(\Gamma,t)}{\tau_{A+1}^A(\Gamma,t)} +
\sum_{i=2,3,..} \frac{n_{A+i}(\Gamma,t)}{\tau_{A+i}^A(\Gamma,t)} ,
\label{eq:EqnA}
\end{equation}
where $\tau_{A+i}^A (\Gamma,t)$ are lifetimes for multiple-nucleon emission with $i \geq 2$.
The two terms in the rhs of Eq.~(\ref{eq:EqnA}) describe the generation of A-nuclei 
in the process with one-nucleon emission $(A+1)+\gamma \to A + N$ and multiple-nucleon emission 
$(A+i)+\gamma \to A + iN$ with $i \geq 2$.  

The exact solution of Eq.~(\ref{eq:EqnA}) can be obtained by our basic method using an 
auxiliary characteristic equation. It reads 
\begin{equation}
n_A(\Gamma,z) = \int_{z}^{z_{max}} \frac{dz'}{(1+z')H(z')}
\left [\frac{n_{A+1}(\Gamma',z')}{\tau_{A+1}^A(\Gamma',z')}+
\sum_{i=2,3,..} \frac{n_{A+i}(\Gamma',z')}{\tau_{A+i}^A(\Gamma',z')}
\right ] \frac{d \Gamma'}{d\Gamma} e^{-\eta_A(\Gamma',z')},
\label{eq:nAexact}
\end{equation}
with 
\begin{equation} 
\frac{d\Gamma'}{d\Gamma} = \frac{1+z'}{1+z} 
exp{\left [ \frac{Z^2}{A} \int_z^{z'} dz'' \frac{(1+z'')^2}{H(z'')}
\left (\frac{\partial b_0^p(\tilde{\Gamma})}{\partial \tilde{\Gamma}} 
\right )_{\tilde{\Gamma}=(1+z'')\Gamma''} \right ] },
\label{eq:dg/dg'}
\end{equation} 
and
\begin{equation} 
\eta_A(\Gamma',z') = \int_z^{z'} \frac{dz''}{(1+z'') H(z'')}
\frac{1}{\tau_A(\Gamma'',z'')} .
\label{eq:etaA1}
\end{equation}
The first generation term in Eq.~(\ref{eq:nAexact}) provides one-nucleon emission
and the second one - the multi-nucleon ($i\geq 2$) emission. We solve 
Eq.~(\ref{eq:nAexact}) by an iteration procedure. In the first iteration we consider only
the terms associated to one-nucleon emission obtaining the already known set of 
densities $n_A^{(1)}(\Gamma,z)$, given by Eq.~(\ref{eq:nA-solut}). 

In the second iteration we include in the injection the sum with multi-nucleon production $i\geq 2$, 
using the densities obtained in the first iteration $n^{(1)}_A(\Gamma,z)$. Solution of 
Eq.~(\ref{eq:nAexact}) gives now a set of densities $n^{(2)}_A(\Gamma,z)$ that can be 
used in the injection term of a new iteration. This iterative procedure rapidly converges, 
because at each new step the multi-nucleon term with $i\geq 2$ adds a small factor being
the lifetime $\tau_{A+i}^A$ with $i\geq 2$ very large. 

We will demonstrate now that already at the second iteration the solution 
approximately coincides with the direct calculation, provided that the 
one-nucleon emission term is much larger than each multi-nucleon 
emission term. 

To solve equation (\ref{eq:EqnA}) we assume that $n_A(\Gamma,t)$ is given by 
\begin{equation}
n_A(\Gamma,t)=n_A^{(1)}(\Gamma,t) + \tilde{n}_A(\Gamma,t),
\label{eq:n1+ntilde}
\end{equation} 
where $n_A^{(1)}(\Gamma,t)$ is the one-nucleon solution  of equation (\ref{eq:kin1}) and 
$\tilde{n}_A(\Gamma,t)$ is a relatively small correction. Putting the sum 
$n_A^{(1)}(\Gamma,t)+\tilde{n}_A(\Gamma,t)$ in equation (\ref{eq:EqnA}), using equation 
(\ref{eq:kin1}) with $Q_A=n_{A+1}^{(1)}/\tau_{A+1}^A$ for one-nucleon emission and 
$n_{A+i}\simeq n_{A+i}^{(1)}$ in the small correction terms with $i=2,3, ...$ in the rhs of 
equation (\ref{eq:EqnA}) one obtains the following equation for $\tilde{n}_A(\Gamma,t)$
\begin{equation}
\frac{\tilde{n}_A(\Gamma,t)}{\partial t} - \frac{\partial}{\partial\Gamma}
\left [ \tilde{n}_A(\Gamma,t) b_A(\Gamma,t)\right ] + 
\frac{\tilde{n}_A(\Gamma,t)}{\tau_A(\Gamma,t)} =
\frac{n_{A+2}^{(1)}}{\tau_{A+2}^A(\Gamma,t)} + 
\frac{n_{A+3}^{(1)}(\Gamma,t)}{\tau_{A+3}^A(\Gamma,t)}+...
\label{eq:Eqntilde}
\end{equation}

The solution of equation (\ref{eq:Eqntilde}) for $\tilde{n}_A(\Gamma,t)$ 
is found by the standard method as 
\begin{equation}
\tilde{n}_A(\Gamma,z) = \int_{z}^{z_{max}} \frac{dz'}{(1+z')H(z')}
\left [\frac{n_{A+2}^{(1)}(\Gamma',z')}{\tau_{A+1}^A(\Gamma',z')}
+ \frac{n_{A+3}^{(1)}(\Gamma',z')}{\tau_{A+3}^A(\Gamma',z')}+...\right]
\frac{d \Gamma'}{d\Gamma} e^{-\eta_A(\Gamma',z')},
\label{eq:nAtilde}
\end{equation}
so that we obtained for $n_A + \tilde{n}_A$ the solution given by 
Eq.~(\ref{eq:nAexact}) with $n_{A+1}=n_{A+1}^{(1)}$ and  
$n_{A+i}=n_{A+i}^{(1)}$ in rhs of this equation,
which coincides exactly with the second-iteration solution. We assume
thus that the second-iteration solution gives an approximate solution 
to the exact equation (\ref{eq:nAexact}), though the strict mathematical 
proof needs the calculations of further iterations.      

In Fig.~\ref{fig5} the total spectrum $n_A(\Gamma,t)$ given by 
Eq.~(\ref{eq:n1+ntilde}) is compared with single-nucleon spectrum 
$n_A^{(1)}(\Gamma,t)$ (the curve below the total). One can see that 
multiple-nucleon emission gives only a small correction. 
\begin{figure}[t!]
\begin{center}
\includegraphics[width=0.46\textwidth,angle=0]{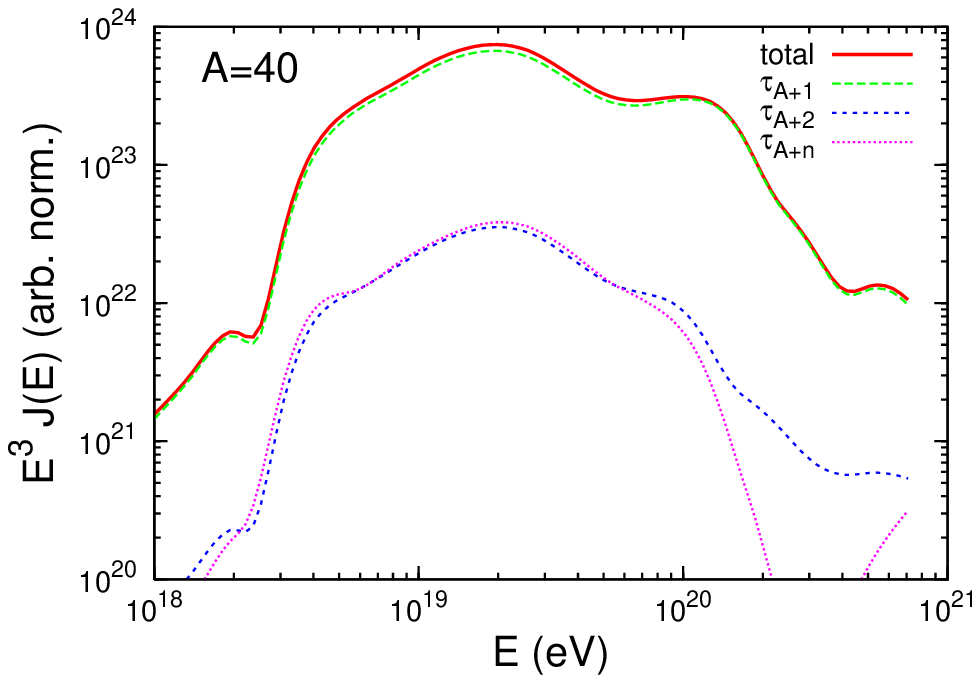}
\includegraphics[width=0.46\textwidth,angle=0]{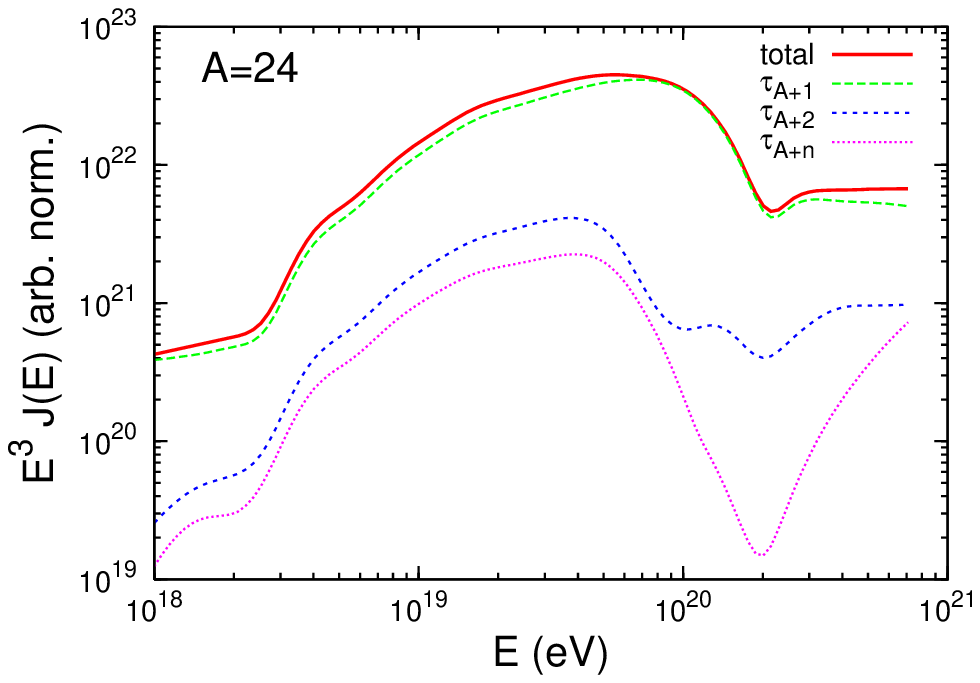}
\caption{Spectra of secondary nuclei $A=40$ (left panel) and $A=24$ 
(right panel) from photo-disintegration of primary Iron. The different   
curves correspond to  photo-disintegration with different multiplicity 
of the emitted protons: the second curve from above (green) for 
multiplicity $i=1$, the third curve for $i=2$ (blue), the lowest 
curve for largest multiplicity $i$ allowed by available experimental 
data, with alpha-particles included. The upper curve (red) gives 
the total flux summed over all available multiplicities.
}
\label{fig5}
\end{center}
\end{figure}

Finally we come over to the calculation of {\em secondary protons}
taking into account the multiple photo-dissociation. 
Kinetic equation for secondary A-associated protons is given by the equation 
\begin{equation}
\frac{\partial n_{A}^p(\Gamma,t)}{\partial t} - \frac{\partial}{\partial \Gamma}
\left [ b_A(\Gamma,t)n_A^p(\Gamma,t) \right ] = Q_p^A(\Gamma,t)
\label{eq:EqnpA}
\end{equation}
where $b_p(\Gamma,t)$ includes adiabatic, pair-production and pion photo-production 
energy losses and $Q_p^A(\Gamma,t)$ is given by 
\begin{equation}
Q_p^A(\Gamma,t)=\frac{n_{A+1}(\Gamma,t)}{\tau_{A+1}^A(\Gamma,t)} + 
2 \frac{n_{A+2}(\Gamma,t)}{\tau_{A+2}^A(\Gamma,t)} + 
3 \frac{n_{A+3}(\Gamma,t)}{\tau_{A+3}^A(\Gamma,t)} + ...
\label{eq:QpA}
\end{equation}

We search for $n_p^A(\Gamma,t)$ solution as the sum of one-nucleon emission 
$n_p^{(1)}(\Gamma,t)$ and two and more nucleon emission $\tilde{n}_p^A(\Gamma,t)$:
\begin{equation}
n_p^A(\Gamma,t) = n_p^{(1)}(\Gamma,t) + \tilde{n}_p^A(\Gamma,t) ,
\label{eq:n_p+tilde}
\end{equation}
where the solutions for two and more nucleon emission are 

\begin{equation}
n_p^{(1)}(\Gamma,z)=\int_z^{z_{max}}\frac{dz'}{(1+z')H(z')}
\frac{n_{A+1}(\Gamma',z')}{\tau_{A+1}^A(\Gamma',z')}
\left (\frac{d\Gamma'}{d\Gamma}\right )_p
\label{eq:npA1}
\end{equation}

\begin{equation}
\tilde{n}_p^A(\Gamma,z)=\int_z^{z_{max}}\frac{dz'}{(1+z')H(z')}
\left [ 2 \frac{n_{A+2}(\Gamma',z')}{\tau_{A+2}^A(\Gamma',z')} +
3 \frac{n_{A+3}(\Gamma',z')}{\tau_{A+3}^A(\Gamma',z')} +... \right ] 
\left (\frac{d\Gamma'}{d\Gamma} \right )_p ,
\label{eq:ntildeAp}
\end{equation}
where $n_{A+i}(\Gamma',z')$ can be taken in one-nucleon or multi-nucleon 
approximation.

For $n_p^A$ given by the sum of the two expressions above, one obtains
\begin{equation}
n_p^A(\Gamma,z)=\int_z^{z_{max}}\frac{dz'}{(1+z')H(z')}
\left [\frac{n_{A+1}(\Gamma',z')}{\tau_{A+1}^A (\Gamma',z')}+ 
2 \frac{n_{A+2}(\Gamma',z')}{\tau_{A+2}^A(\Gamma',z')} +
3 \frac{n_{A+3}(\Gamma',z')}{\tau_{A+3}^A(\Gamma',z')} +... \right ] 
\left (\frac{d\Gamma'}{d\Gamma} \right )_p ,
\label{eq:npAtot}
\end{equation}

Corrections due to multiple-nucleon photo-dissociation
in Eq.~(\ref{eq:npAtot}) for protons are larger than for nuclei A,
because of the coefficients 2,~3,~ etc imposed by proton multiplicity.
Nevertheless, these corrections at the energies relevant for UHECR physics 
are still very small (see the discussion and calculations below). 
\begin{figure}[t!]
\begin{center}
\includegraphics[width=0.9\textwidth,angle=0]{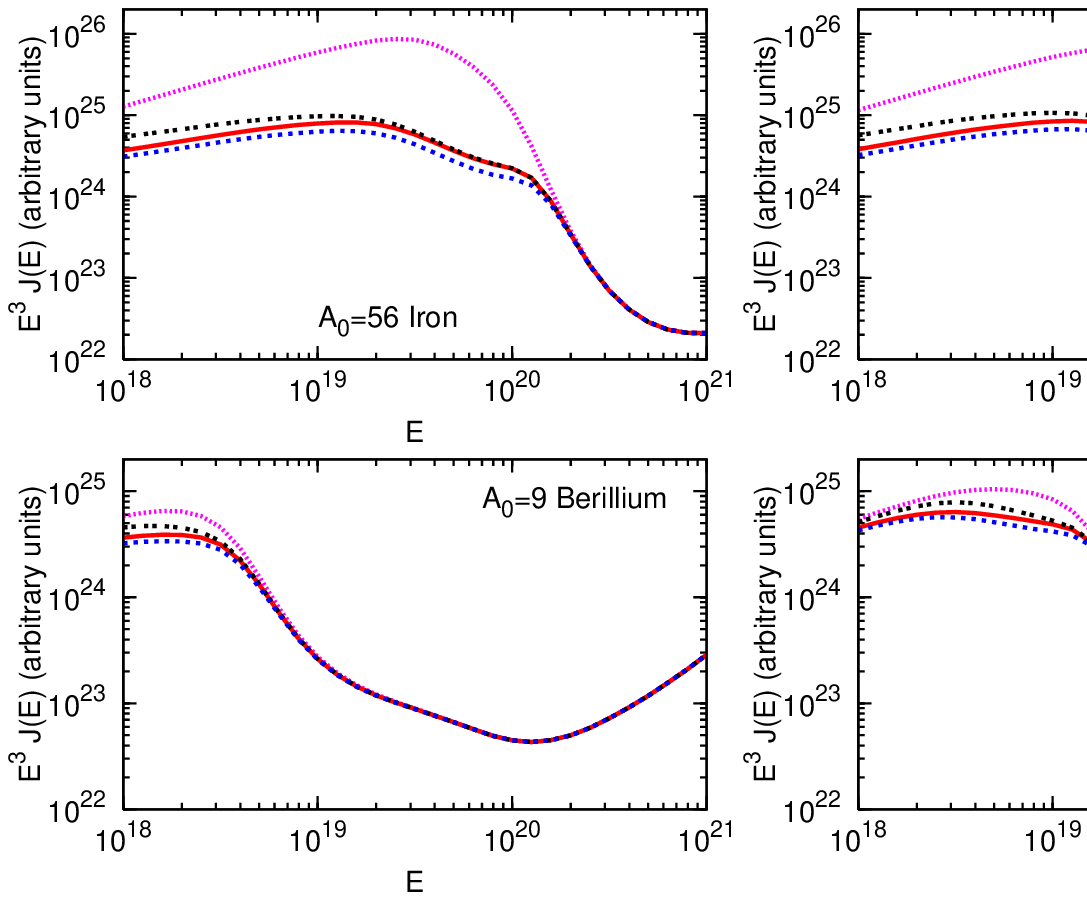}
\caption{Flux of the primary nuclei for different injected nuclei species 
$A_0$ (as labelled). Three different choices 
of EBL are plotted: baseline model (red continuous), fast evolution 
(blue dotted) and minimum EBL (black dotted).
The case of CMB alone (see paper I) is plotted by dotted magenta curve.
The spectra of primary nuclei are always suppressed by EBL. The
largest suppression corresponds to the fast-evolution model with 
the highest EBL flux at larger redshifts.
}
\label{fig6}
\end{center}
\end{figure}

\section{Spectra}
\label{sec:spectra}

In this section we will discuss the spectra of different species of 
primaries, as well as the spectra 
of secondary nuclei and protons, produced by 
photo-disintegration of the primaries during propagation. 
The main emphasis is given to the impact of 
the EBL on the spectra. We will consider the three models for the EBL 
cosmological evolution discussed in section \ref{sec:losses}: 
the baseline  and the fast evolution
models of \cite{stecker06}, and the minimum EBL as presented in 
section \ref{sec:losses}. At $z=0$ all three EBL models are 
normalized to the observed photon spectrum, they differ only 
at larger redshifts with the fast-evolution model giving the largest 
flux. In order to isolate 
the impact of the background radiation on the spectra, we will 
assume that only one nucleus specie $A_0$ is accelerated and 
injected into space according to Eq.~(\ref{eq:injA0}). The general 
case of mixed injection composition
can be obtained by summing the fluxes with different $A_0$. 
Sources are assumed homogeneously distributed in the universe. 

The injection spectrum is taken in a power-law form with the 
power-law index and maximum energy at the source 
fixed as $\gamma_g=2.3$ and $E_{\max}=Z_0\times 10^{21}$ eV.
The emissivity ${\mathcal L}_0$, i.e. the energy injected per unit
of comoving volume and per unit time, is not specified and the calculated 
diffuse spectra are given in arbitrary units.

\subsection{Primary nuclei}
\label{sec:primary-nuclei}

The spectra of primary nuclei are the simplest for calculation and 
understanding. The spectrum of primary $A_0$ is given by 
Eq.~(\ref{eq:nA0-solut}) with the generation term $Q_{A_0}$ 
and survival probability $\exp(-\eta_0)$ described by 
Eq.~(\ref{eq:injA0}) and Eq.~(\ref{eq:etaA}), respectively. 
Because of the large space density of the CMB 
photons, the spectrum of primary nuclei is formed first due 
to interaction with the CMB 
photons, and then is distorted (much more slowly) by the EBL. As
explained above, as far as the EBL is concerned, only 
photo-disintegration is important 
and should be taken into account. The primary nuclei 
are photo-disintegrated by the EBL photons only at low 
Lorentz factors (at $\Gamma \leq 2\times 10^9$, see section
\ref{sec:losses}) i.e. at these Lorentz
factors the primary spectrum is depleted by the EBL. Numerically 
this depletion 
is described by diminishing of $\tau_{A_0}$ and thus by decreasing of 
survival probability $\exp(-\eta_{A_0})$, which suppresses the flux 
(\ref{eq:nA0-solut}). A side effect of this interaction is the
production of secondary nuclei and protons with the same Lorentz-factors 
$\Gamma \leq 2\times 10^9$. Fig.~\ref{fig6} confirms these
expectations. The spectra calculated with CMB alone are shown by
dotted magenta curves. The EBL radiation {\em always suppresses} these
spectra. All three EBL models (baseline, fast-evolution and minimum 
EBL) show almost identical suppression, because they have equal photon 
fluxes at $z=0$ and differ only at larger $z$ due to evolution, with fast
evolution model giving the largest photon flux. For heavy nuclei the
suppression is stronger. At large energies, where CMB dominates,
suppressions are equal. 

\begin{figure}[t!]
\begin{center}
\includegraphics[width=0.9\textwidth,angle=0]{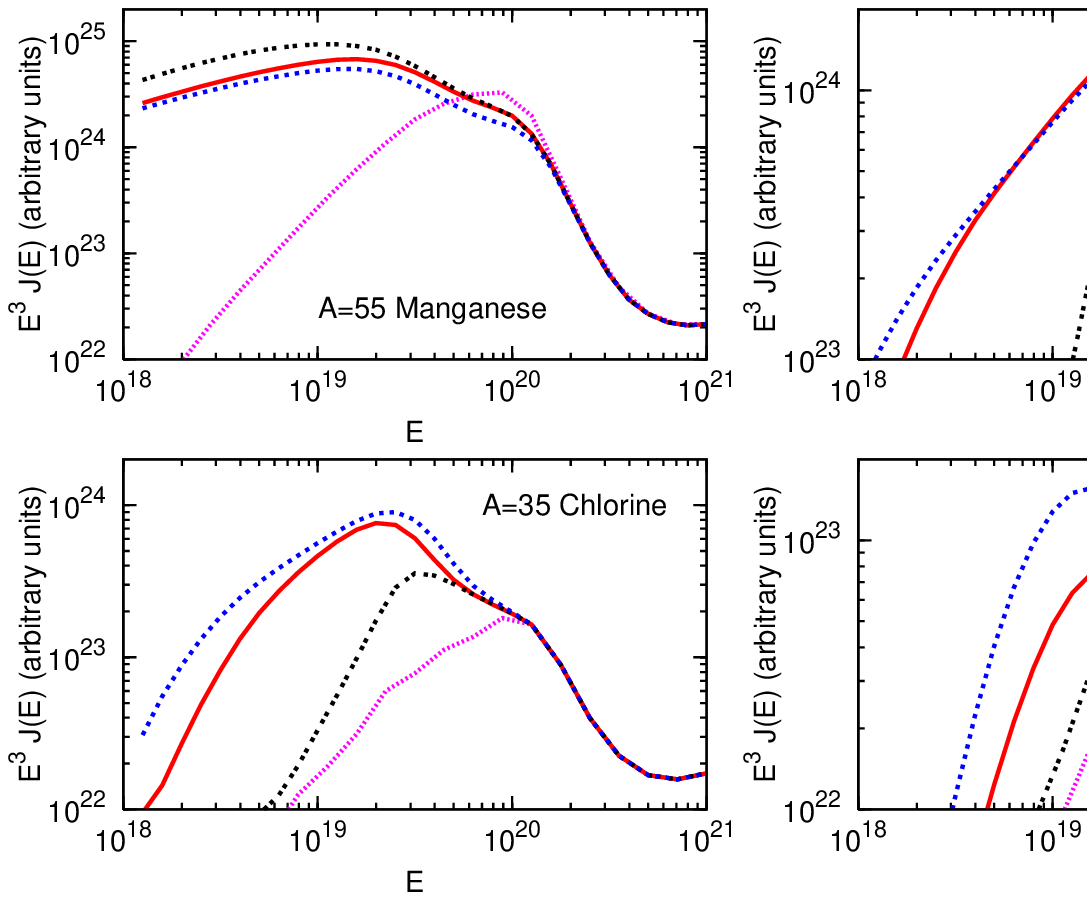}
\caption{Spectra of heavy secondary nuclei (as labelled) produced in the 
photo-disintegration chain of Iron.
Three models with different evolution of the EBL are shown: baseline model 
(red continuous), fast-evolution (blue dotted) and minimum EBL (black dotted). 
The case of CMB alone (see paper I) is plotted by dotted magenta curve.
In contrast to primary nuclei, EBL increases the flux of secondary nuclei
at lower energies: One can see the low-energy tails in the spectra, produced 
by EBL (see text for explanation).
}
\label{fig7}
\end{center}
\end{figure}

\subsection{Secondary nuclei}
\label{sec:secondary-nuclei}

In the dominant single-nucleon approximation the spectrum of secondary 
nuclei is given by Eq.~(\ref{eq:nA-solut}), with generation rate and 
survival probability presented by Eqs.~(\ref{eq:injA-single}) and 
(\ref{eq:etaA}), respectively. In presence of the EBL these two factors work 
in opposite directions: the survival probability in case of small $\tau_A$ 
suppresses the flux just like in the case of primary nuclei, while the generation 
term, inversely proportional to $\tau_{A+1}$, increases the flux. We can give a 
qualitative argument, based on the trajectory calculations, proving that the 
generation term {\em dominates} and interaction with EBL radiation 
{\em always increases} the secondary nuclei flux in comparison with the 
case of CMB alone (see Figs. \ref{fig7},~\ref{fig8}).

Consider first the case of the CMB only and a secondary nuclei $A$ 
at $z_0=0$. Let us study like in section 2.2 of paper I the backward 
evolution trajectories $A(z)=\mathcal{A}(A,\Gamma,z)$ and 
$\Gamma(z)=\mathcal{G}(A,\Gamma,z)$, where $A$ and $\Gamma$ are the
values at $z_0=0$. At the generation red-shift
$z=z_g$, by definition $A(z_g)=A_0$, and $\Gamma(z_g)=\Gamma_g$. 
Let us now switch on the EBL. $A(z)$ starts to increase earlier and
reaches $A_0$ earlier, i.e. at smaller $z_g$. Hence 
$\Gamma_g=\Gamma (z_g)$ decreases, 
and the number of generated primary nuclei 
$Q_{A_0} \propto \Gamma_g^{-\gamma_g}$ increases. 

We may put this effect in other words: 
the EBL accelerates the evolution of $A(z)$, and $z_g$, where $A(z)$ reaches 
$A_0$, becomes smaller, $\Gamma_g$ becomes smaller, too, and the generated
flux becomes larger. 

The fluxes of secondary nuclei are shown in Fig.~\ref{fig7} for heavy
nuclei and in Fig.~\ref{fig8} for light nuclei. The fluxes computed with 
CMB only are shown by magenta dotted lines. As anticipated these fluxes
are always lower than ones computed with EBL taken into account (three
upper curves). The fluxes corresponding to different EBL also obeys 
the above hierarchy: the stronger EBL, the larger secondary-nuclei
flux. In Figs.~\ref{fig7} and \ref{fig8} one can see that minimum EBL 
(black dotted curves) corresponds to the lowest flux among the three 
EBL versions. The only exception is given by $A=55$, because of its
"vicinity" to the primary injected $A=56$.  

\begin{figure}[t!]
\begin{center}
\includegraphics[width=0.9\textwidth,angle=0]{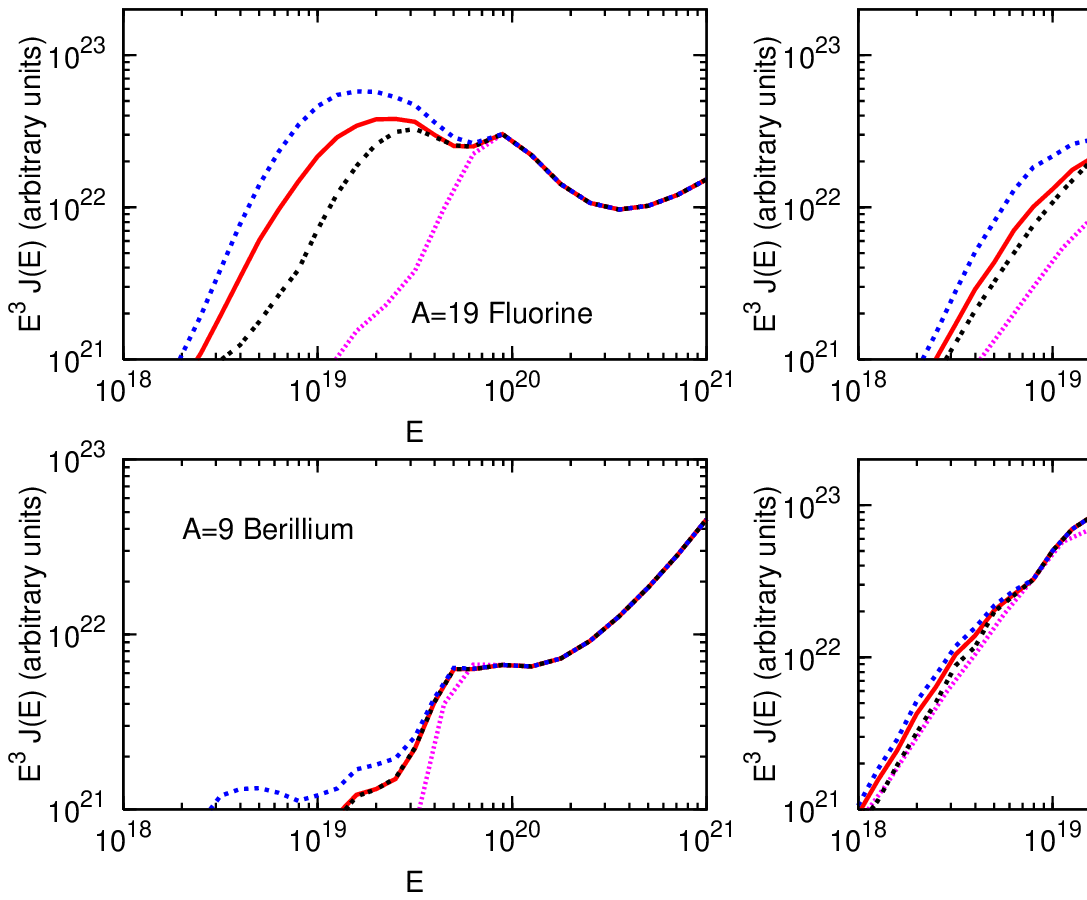}
\caption{The same as  Fig.~\ref{fig7} for light secondaries.
The largest EBL flux is given in the fast-evolution model. 
}
\label{fig8}
\end{center}
\end{figure}

\subsection{Secondary protons}
\label{sec:secondary-protons}

The flux of secondary protons $p$ born simultaneously with the
brother-nuclei $A$ in the process $(A+1)+\gamma \to A+N$ is given 
by Eq.~(\ref{eq:np}), with the same generation rate (\ref{eq:injA-single})
as for nuclei $A$. We call these protons $A$-associate and denote 
their space density $n_p^A(\Gamma,z)$. The total proton flux (density) 
is calculated summing up $n_p^A(\Gamma,z)$ over all $A < A_0$. In this  
section we limit ourselves by $A_0=56$.

As one can see from Eq.~(\ref{eq:np}) the flux of secondary protons 
is affected by the EBL only through the injection term $Q^A_p$, 
i.e. through the photo-disintegration lifetime of the parent 
nucleus $A+1$ (see Eq.~\ref{eq:tauA-single}). 
Increasing the EBL flux results in decreasing the lifetime 
$\tau_{A+1}$, making more efficient the production of secondary protons.
One can see this correlation of the proton fluxes in Figs.~\ref{fig9} and
\ref{fig10}: the proton fluxes increase with the EBL flux.
This effect is restricted to the Lorentz factor range 
$1\times 10^8< \Gamma < 2\times 10^9$, where the effect of the EBL
plays a relevant role (see figures \ref{fig1} and \ref{fig2}). 
Caused by the Lorentz-factor equality $\Gamma_{A+1}=\Gamma_p$, it
corresponds to dominant proton production on the EBL with energies 
$E_p \le 2\times 10^{18} $ eV. For larger energies $E_p$ and larger 
red-shifts the production on the CMB dominates 
(see Figs.~\ref{fig9} and \ref{fig10}).  
From this discussion it follows that the proton production on the EBL 
occurs in the energy range of minor importance for UHECR study. 

\begin{figure}[t!]
\begin{center}
\includegraphics[width=0.9\textwidth,angle=0]{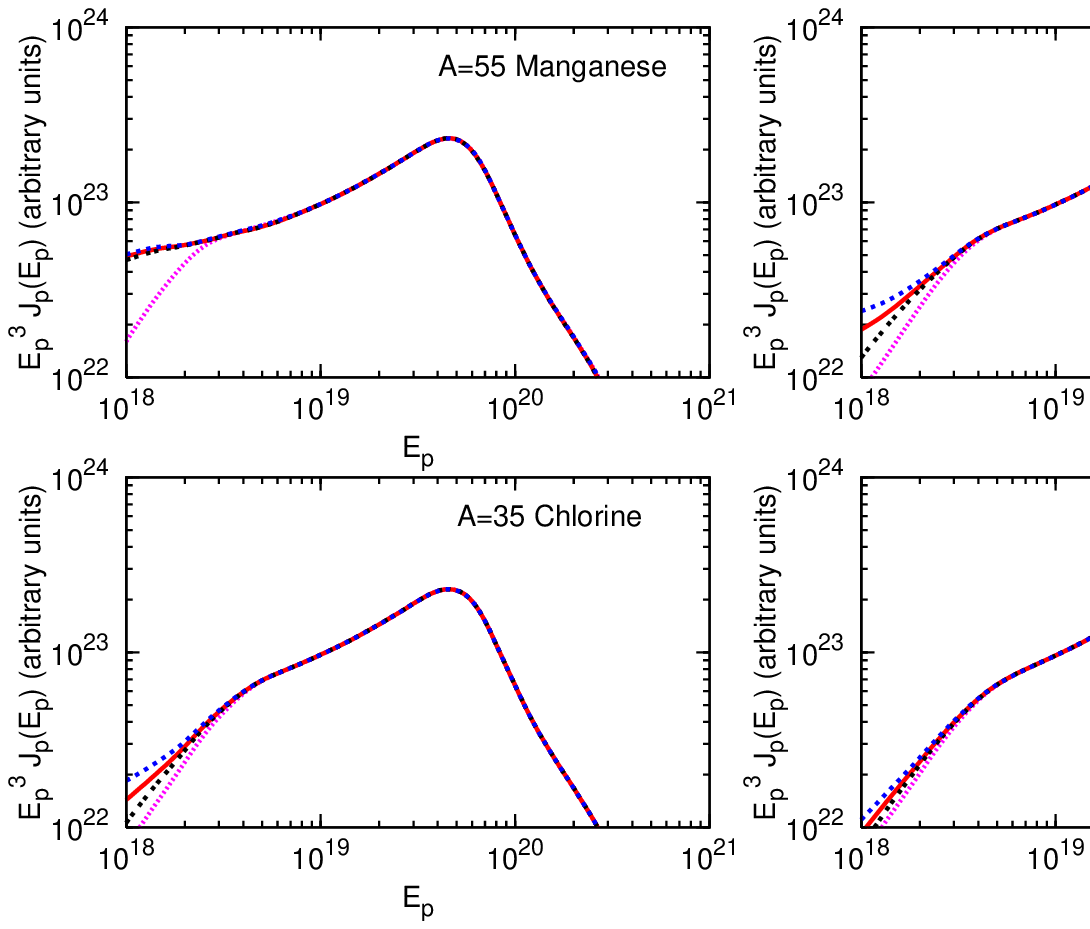}
\caption{Flux of secondary protons accompanying production of secondary 
$A$-nuclei (as labelled). Three different versions of the 
EBL are plotted: baseline model (red continuous), fast evolution (blue
dotted) and minimum EBL (black dotted).
The case of CMB alone (see paper I) is shown by dotted magenta curve.
One may observe the correllation of proton fluxes with EBL: at energy 
$E_p \sim (1 - 3)\times 10^{18}$~eV the larger proton flux
corresponds to larger EBL (see text).
}
\label{fig9}
\end{center}
\end{figure}
\begin{figure}[h!]
\begin{center}
\includegraphics[width=0.9\textwidth,angle=0]{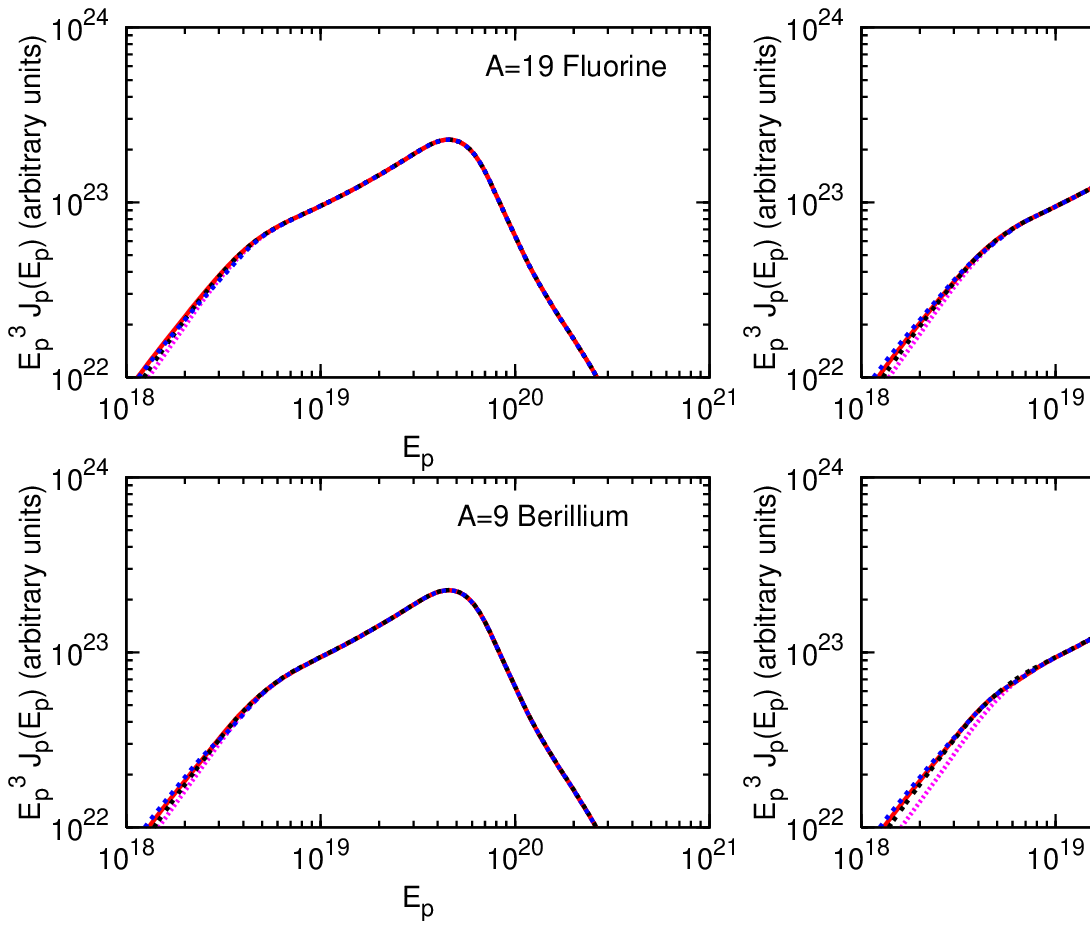}
\caption{The same as in figure \ref{fig9} for lighter secondaries (as labelled).}
\label{fig10}
\end{center}
\end{figure}

\subsection {Secondary nuclei and protons from primary Iron}
\label{sec:pureFe}

To conclude Section \ref{sec:spectra} we expose in Fig.~\ref{fig11} the
predicted fluxes of secondary nuclei and secondary protons produced by 
the primary nuclei with a pure iron composition at the source.  The
injection parameters are fixed as in Section \ref{sec:CKE}. The fluxes
of secondary nuclei are grouped summing over different nuclei species, 
as shown in the figure, while the flux of secondary protons is given as 
the total flux summed over all associated nuclei with $A \le A_0$. 
Four groups of nuclei are presented in Fig.~\ref{fig11}: primary Iron, 
heavy mass secondaries $40 < A < 56$, intermediate mass secondaries 
$26 < A < 39$ and light mass secondaries $2 < A < 25$. One can observe that 
primary Iron and heavy secondaries dominate over secondary protons and    
light secondary nuclei. 

\begin{figure}[t!]
\begin{center}
\includegraphics[width=0.46\textwidth,angle=0]{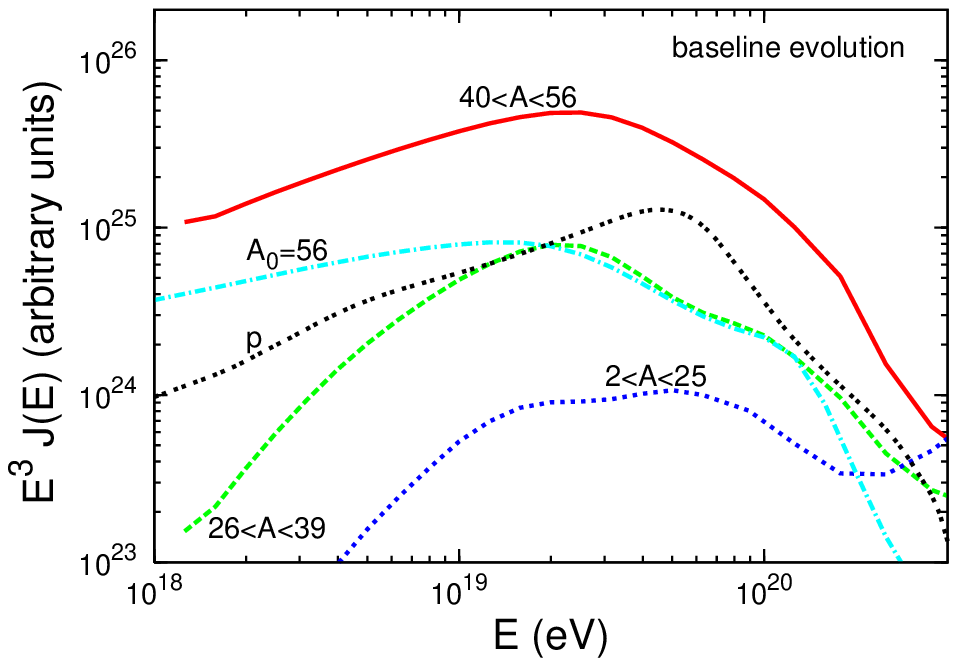}
\includegraphics[width=0.46\textwidth,angle=0]{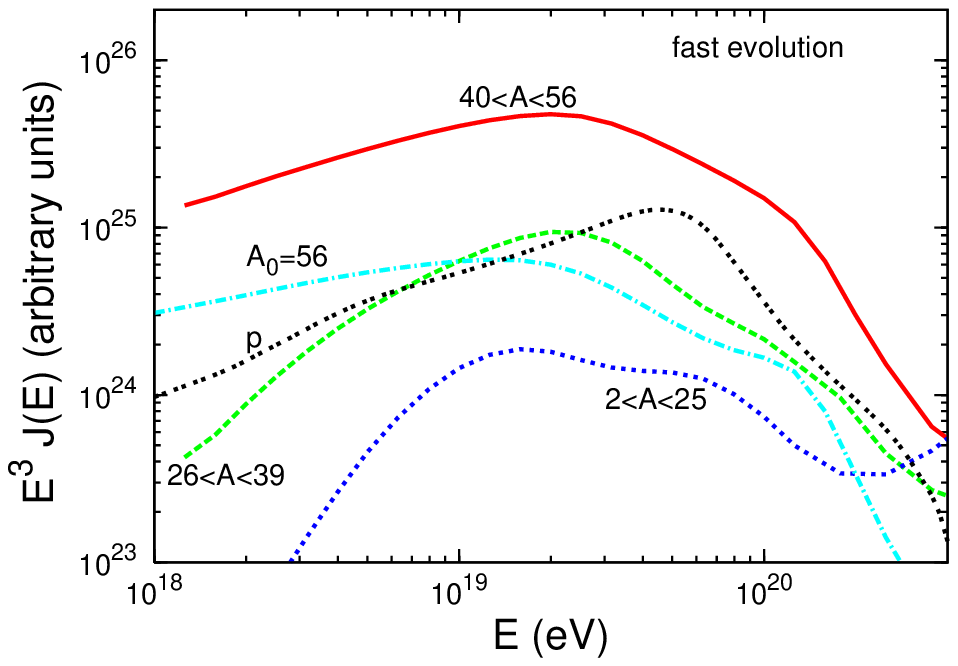}
\caption{Fluxes of secondary nuclei and protons produced by 
the injection of pure Iron at the source with  injection parameters 
as in section \ref{sec:CKE}. EBL fluxes are taken according   
to baseline model (left panel) and fast-evolution model (right
panel). The secondary-nuclei fluxes are summed over $A$ as labelled.
The secondary proton flux is summed over all $A$ and is shown
by black dotted curve. 
}
\label{fig11}
\end{center}
\end{figure}

EBL is responsible for the production of secondary protons with energy 
$E \lsim (1 - 2)\times 10^{18}$~eV, which are not of much  
interest for UHECR.  As one can see from Figs. \ref{fig1} and 
\ref{fig2}, EBL dominates nuclei energy losses at Lorentz-factors 
$\Gamma \lsim (1 - 2)\times 10^9$. Therefore protons are
produced with the same Lorentz-factors, i.e. with energies below 
$(1 - 2)\times 10^{18}$~eV. In the production of protons with higher 
energies CMB radiation strongly dominates. This fact explains also 
the suppression of multiple-proton production in photo-disintegration.  
In the case of CMB the photo-disintegration at intermediate energies 
occurs at the threshold, where single nucleon production strongly 
dominates, while at very high energies primary nuclei are
photo-disintegrated almost simultaneously, and using single or 
multiple nucleon production makes no difference.
In the case of EBL multiple photo-disintegration might be important 
but as was discussed above protons are produced at too low 
energies for UHECR.   

Light mass secondary nuclei follow the general rule of an increasing
flux with increasing EBL (compare the right and left panels 
with higher and lower flux of EBL, respectively). The difference is
seen for $2 < A <25$ secondary nuclei. Suppression of Iron flux 
(primary nuclei) in the case of larger EBL flux is seen at lower and 
intermediate energies. For secondaries with $26 < A <39$ the increase 
of flux with stronger EBL (right panel) is also present. For secondary
protons and heavy mass secondary nuclei ($40 < A < 56$) the correlation with 
EBL is weak in accordance with Fig.~\ref{fig8} and Fig.~\ref{fig10}}.

\section{Comparison with other computation schemes}
\label{sec:comparison}

As discussed in the Introduction to paper I the study of the
propagation of UHE nuclei in astrophysical backgrounds has recently 
attracted much attention because of the experimental results of the 
Auger Observatory. Different computation schemes were presented 
in literature based on both analytic and MC approaches. In this subsection we
will briefly compare our results with those obtained by other methods. 

\begin{figure}[h!]
\begin{center}
\includegraphics[width=0.5\textwidth,angle=0]{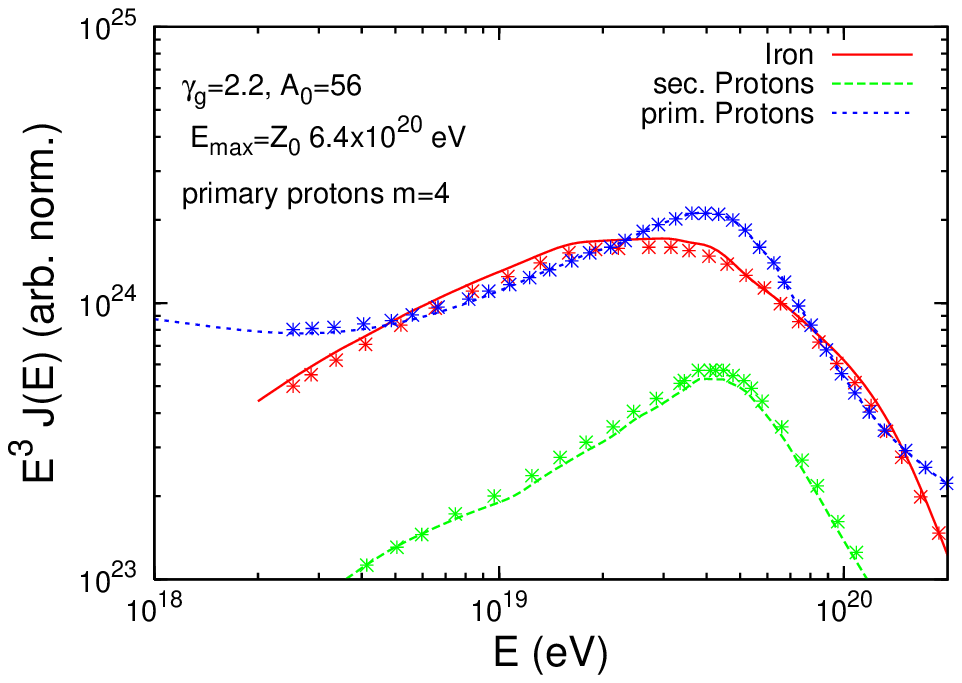}
\caption{Flux of primary protons, secondary protons and Iron as computed in 
\cite{Kalashev1} (asterisks) and in our calculations (lines). The injection 
power-law index is $\gamma_g=2.2$ with a maximum acceleration energy for protons 
$E_{\max}=6.4\times 10^{20}$. The injection of primary protons is assumed with a 
strong cosmological evolution $m=4$ (see text).}
\label{fig12}
\end{center}
\end{figure} 

The analytic calculations presented in literature are all based on 
the kinetic equations, differences among different analytical approaches 
are all related to the collision term used in such equations. Io our discussion
here we follow the approach and definitions of \cite{Pitaevskii}. 
Moreover, we can avoid the analysis connected with magnetic fields 
because for a homogeneous distribution of sources the energy spectra do not
depend on magnetic fields (propagation theorem \cite{prop-theor}).  
The most noticeable works based on the kinetic-equation methods are  
\cite{Kalashev1} - \cite{Taylor2}. 

In \cite{Taylor1,Taylor2} a Boltzmann type collision term is used and the 
analytic solution of the kinetic equation is determined through a discretization 
procedure. We cannot  compare accurately our solutions with 
those of \cite{Taylor1,Taylor2} because of difference in equations and 
lacking of some numerical parameters, such as the maximum and 
minimum red-shift assumed in the source distribution of \cite{Taylor1}. 
Moreover, in \cite{Taylor1,Taylor2} the authors use a different model for 
the EBL evolution \cite{Franceschini} respect to the model used here. 

Comparison with the results of \cite{Kalashev1,Kalashev2} 
has a crucial importance for us because in these works the same kinetic equations
and EBL are used, and numerical parameters are not much different respect to ours. 
The most important difference among our approach and that of 
\cite{Kalashev1,Kalashev2} is that kinetic (transport) equations are solved using 
a numerical procedure (see \cite{Kalashev2}) and thus these computations provide 
a numerical test of our analytical method.
      
In Fig.~\ref{fig12} we compare our results with those from 
figures 3 and 6 of \cite{Kalashev1}, in particular we plot the case of the injection 
of primary Iron nuclei and protons with a power-law index $\gamma_g=2.2$ and a 
maximum energy for protons $E_{\max}=6.4\times 10^{20}$ eV. The injection 
of primary protons in \cite{Kalashev1} is assumed with a strong cosmological 
evolution: $Q_p(\Gamma,z)\propto (1+z)^m$ with $m=4$ \cite{dip}. From 
Fig.~\ref{fig12} one can see a very good agreement of our results
with those of \cite{Kalashev1} (the curves present our calculations and 
asterisks - numerical calculations from \cite{Kalashev1}). 

A completely different approach to solve the problem of the propagation of UHE 
particles in astrophysical backgrounds is provided by the MC technique. In this case
there is a reach literature with different MC schemes presented (see the discussion
and references in the introduction to paper I). In this section we will present a comparison 
of our results with those of two different MC computations
\cite{Allard}, \cite{Sigl1} - \cite{Sigl2}.
In these two cases the authors consider the injection of pure Iron with a power law index 
of the injected particles $\gamma_g=2.3$ and a maximum generation energy of the iron 
nuclei $E_{\max}^{\rm Fe}=5\times 10^{21}$ eV. The EBL radiation assumed in this 
comparison is the baseline model of \cite{stecker06}, which we use too.

\begin{figure}[t!]
\begin{center}
\includegraphics[width=0.46\textwidth,angle=0]{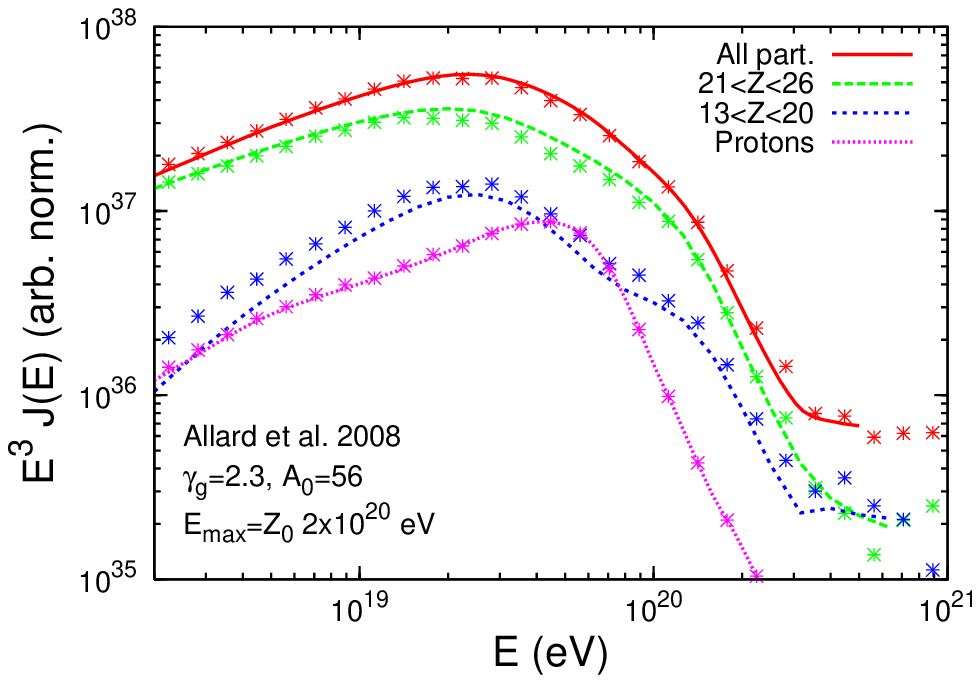}
\includegraphics[width=0.46\textwidth,angle=0]{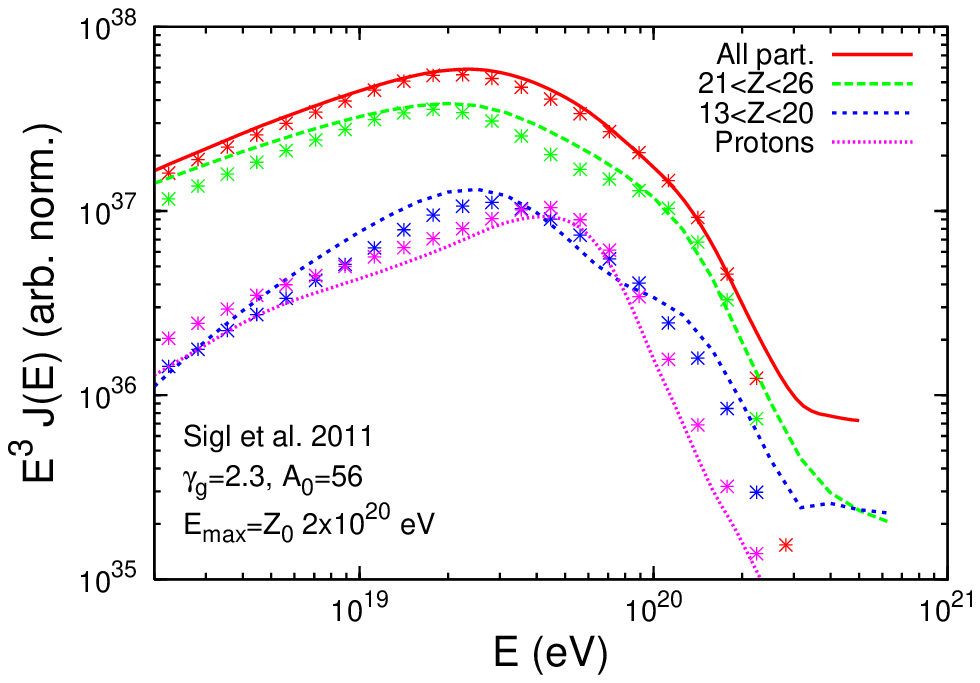}
\caption{Fluxes of secondary nuclei and protons produced by the injection of
pure Iron at the source as obtained by MC computation (asterisks) and in our 
kinetic-equation  approach (lines). The spectra are calculated for all particles 
(primary Iron, all secondary nuclei and secondary protons) shown by red,
for heavy and intermedium secondary nuclei (green and blue) and for secondary protons  
(majenta). In the left panel the MC spectra are from
\cite{Allard} and in the right  panel from \cite{Sigl1} - \cite{Sigl2}.
}
\label{fig13}
\end{center}
\end{figure}
 
Figure \ref{fig13} shows a good agreement of these computations with ours. 
The comparison includes the total flux of primary and secondary nuclei summed with  
the flux of secondary protons. Shown also in this figure are the flux of secondary protons 
and the secondary nuclei grouped into $21\le Z \le 26$ (heavy mass group) and
$13\le Z \le 20$ (intermediate mass group). The lighter nuclei are also calculated in both MC 
and our works. They have low fluxes, less that a few percent of the total flux. 
The same EBL taken from \cite{stecker06} is used in all calculations presented in 
Fig.~\ref{fig13}. The fluxes from MC simulations are taken from figure 5
of \cite{Allard} (left panel) and from \cite{Sigl1} - \cite{Sigl2} (right panel).  
The spectra from our analytic calculations are shown by lines and from
MC simulations - by asterisks. The agreement is good except the highest energies in the 
right panel. We use there the data from \cite{crpropa2,OurMC} where the highest energy recovery 
is absent in contrast to \cite{Sigl2} where this recovery is present (see Fig.~7 of \cite{Sigl2}). 
As clearly explained in \cite{Sigl2} the spectrum recovery in this simulation depends on the 
nearby source location relative to the observer. The uncertainties given by cosmic variance is 
very large at the highest energies (see Fig.~7 of \cite{Sigl2}) and can include a strong recovery of
the spectrum or its absence. In our calculations with a homogeneous distribution of sources 
the recovery of the spectrum at the highest energies is a compulsory feature.

The agreement of our kinetic approach with the MC computations is quite good. As expected, 
the fluxes of secondary nuclei with large atomic mass $A$, close to the primary injected $A_0$ in the 
photo-disintegration chain, show a better agreement than the fluxes of secondaries 
with lower atomic mass number. The latter undergo a larger number of photo-disintegrations 
collisions and therefore the computation of their flux is more affected
by uncertainties in both MC and kinetic-equation approaches. 

As follows from the comparison of figures \ref{fig13}, the total 
flux of nuclei expected on earth is less affected by these uncertainties 
because it is dominated by the flux of nuclei with the larger atomic mass 
number (see also discussion in the next section). 

\section{Discussion and Conclusions}
\label{sec:conclusions}

The present paper is the continuation of the accompanying paper I,
where different methods for the analytic calculations of UHE nuclei 
spectra have been studied, including CMB as the only background 
radiation.The exact knowledge of the CMB spectrum at any red-shift
provides a clear understanding  of the nuclei spectrum-shape and its 
production. In the present work we have performed more realistic 
calculations, including additionally the EBL as background radiation 
and focusing on its role. We used the Coupled Kinetic Equations (CKE) 
method as the most transparent and precise. The most important element 
of this method is the generation rate common for the production of 
secondary nuclei $A$ and secondary nucleons $N$ in the process 
$(A+1)+\gamma_{\rm bcgr} \to A+N$. For single-nucleon 
photo-disintegration it has a form 
$n_{A+1}(\Gamma)/\tau_{A+1}^A(\Gamma)$ and is given explicitly by 
Eq.~({\ref{eq:injA-single}). It couples two successive kinetic equations 
for $n_A(\Gamma)$ and $n_{A+1}(\Gamma)$ and has a key importance 
for the calculation of $n_A(\Gamma)$ and $n_p(\Gamma)$. 

In the calculations we included also the multiple-nucleon 
photo-disintegration.

In this section we discuss the impact of the EBL on the calculated 
spectra in comparison with CMB. The physical understanding of these
effects is a great advantage of the analytic methods, which we use.  

The role of the EBL is limited by photo-disintegration of the
low-energy nuclei with Lorentz-factors in the range 
$1\times 10^8 < \Gamma < 2\times 10^{9}$, when the energies of CMB 
photons become too low for photo-disintegration. At the same time 
the continuous $e^+e^-$-energy loss occurs in this energy range  
due to the interaction with CMB photons as much more numerous. At 
$\Gamma > 2\times 10^9$ both photo-disintegration and pair production
are dominated by CMB and the spectra calculated in paper I become fully 
applicable.  

EBL distorts the nuclei spectra in two-fold way. On one hand,
destroying the nuclei  this radiation suppresses the nuclei flux. 
On the other hand in this process it produces lighter secondary 
nuclei, increasing their flux.

Let us discuss first the impact of the EBL on the primary nuclei $A_0$
accelerated in the sources. Their spectrum is formed first due to the
interaction with CMB, because the number of CMB photons is much  
larger than that of EBL. The role of the EBL consists in a flux
suppression in the range $1\times 10^8 < \Gamma < 2\times 10^{9}$
due to photo-disintegration. The flux of primary nuclei $A_0$ 
diminishes with increasing EBL.  The calculated spectra of primary 
nuclei are exposed in Fig.~\ref{fig6}. The EBL suppression of the 
spectra is clearly seen. The steepening of spectra is much different 
from GZK cutoff.

Consider next the secondary nuclei $A$ in the same Lorentz 
factor interval as above. Apart from the described 
suppression of the flux, there is its regeneration due to the 
photo-disintegration of  $A+1$ (or heavier) nuclei on the EBL. As 
it is explained in section~\ref{sec:secondary-nuclei} this
regeneration always dominates and thus increasing EBL flux results 
in increasing the secondary nuclei flux. As a result in CMB-produced 
$A$-nuclei spectrum (see Figs.~\ref{fig7} and \ref{fig8}) the 
low-energy tail appears. This effect is most remarkable influence of 
the EBL photo-dissociation on the CMB-produced spectrum. One can clearly 
see it in Figs.~\ref{fig7} and~ \ref{fig8}. In section 
\ref{sec:manynucleon} we developed the method of including 
multiple-nucleon photo-disintegration in the CKE method of calculation of
spectra of secondary nuclei and protons. 

Finally, we discuss secondary protons. They are produced in 
$(A+i)+\gamma_{\rm bcgr} \to A+ iN$ photo-dissociation with the same 
Lorentz-factor as the accompanying $A$-nucleus and with the same
generation rate. The total density of the secondary protons is found by 
summing $n_p^A(\Gamma)$ over all $A < A_0$. The protons are
produced in the same Lorentz-factor range as the secondary nuclei,
i.e. with energies $1\times 10^{17} - 2\times 10^{18}$~eV due to
interaction with EBL.  In contrast to the parent nuclei these 
protons are not photo-disintegrated and undergo small energy losses. 
At higher energies protons are produced on CMB from nuclei with larger
Lorentz factors. 

In (unrealistic) models where only heavy nuclei, e.g. Iron, are 
accelerated at the sources, photo-disintegration can be the only 
mechanism for proton production. It can comprise about $10\%$ 
of the total flux with maximum ratio reaching $20\%$ at 
$5\times 10^{19}$~eV (see Fig.~\ref{fig11}). These values depend on 
$E_{\max}$, $\gamma_g$ and $A_0$. taken as $2.6\times 10^{22}$~eV, 
2.3 and 56, respectively. For harder spectrum with $\gamma_g=2.1$ 
the maximum ratio becomes $25\%$ and it diminishes for lower 
$E_{\max}$ and $A_0$. In principle the above ratios can be 
understood as lower limits, because in more realistic models a directly 
accelerated proton component is always expected. 

However, depending on the maximum energy, there could be models 
with zero proton fraction at the highest energies. An example is given 
by the 'disappointing model' \cite{third}, where the maximum acceleration 
energy for protons is fixed at $3\div 4 \times 10^{18}$ eV. In this case 
primary and secondary protons are absent in the energy interval between 
$E_p^{\max}$ and $Z_0 E_p^{\max}$, since the energy of secondary  
protons do not exceed $(Z_0/A_0) E_p^{\max} \simeq 0.5 E_p^{max}$.

One can see from all calculated spectra that the differences in the cosmological 
evolution regimes for the EBL do not change strongly the spectra. It 
occurs because all evolution regimes are normalized by the same
EBL flux at $z=0$, while CMB radiation becomes more essential at large 
$z$ due to a more rapid evolution. An exceptional case is given by the
spectra of secondary nuclei, see Figs.~\ref{fig7} and \ref{fig8}, because 
all the effect on the low-energy tail is caused by the EBL. 

In Fig.~\ref{fig11} the fluxes of secondary nuclei and protons 
produced by primary (accelerated) Iron nuclei are exposed. The
secondary nuclei are presented as three groups: heavy secondaries 
(summed over $A$ from 40 to 56), the intermediate group 
($A$ from 26 to 39) and the light group ($A$ from 2 to 25).  
The proton flux is summed over all $A$. One can see that heavy
secondary nuclei dominate. The spectrum 'cutoff' caused by
photo-disintegration is less steep and starts earlier than the GZK 
cutoff. Flux of secondary protons is subdominant. 

As the main result of these two papers we consider the study of the
fundamental properties of propagation of UHE nuclei through CMB and 
EBL in an analytic approach. In a forthcoming paper of this series 
we will build more realistic models for a direct comparison with the 
observational data.

\section*{Acknowledgements}
We thank Pasquale Blasi, Yurii Eroshenko and Askhat Gazizov for valuable
discussions. This work is partially funded by the contract
ASI-INAF I/088/06/0 for theoretical studies in High Energy
Astrophysics and by the Gran Sasso Center for Astroparticle Physics (CFA)
funded by European Union and Regione Abruzzo under the contract P.O. FSE
Abruzzo 2007-2013, Ob. CRO. The work of SG is additionally  funded by 
the grant of President of RF SS-3517.2010.2, VB and SG - by FASI grant 
under state contract  02.740.11.5092.

\end{document}